\documentclass[preprint]{aastex}
\pdfoutput=0

\usepackage{amsmath}
\usepackage{longtable}
\usepackage{natbib}

\usepackage{graphicx}
\usepackage{subfig}
\usepackage[figuresright]{rotating}

\newlength{\FigureWidth}
\usepackage{subfig}
\newlength{\CaptionWidth}
\newlength{\SubFloatSpace}
\newlength{\CaptionSpace}
\usepackage[
colorlinks=true
]
{hyperref}
\usepackage[all]{hypcap}
\hypersetup{
pdfpagemode = {UseOutlines},
baseurl = {http://www.mpia.de/homes/kuiper},
pdftitle = {Circumventing the radiation pressure barrier in the formation of massive stars via disk accretion},
pdfauthor = {R.~Kuiper},
pdfkeywords = {}
}

\newcommand{\vONE}[1]{#1}
\newcommand{\Msol}{\mbox{$\mbox{ M}_\odot$} }

\newcommand{\rhocgs}{\mbox{$\mbox{ g} \mbox{ cm}^{-3}$} }

\shorttitle{Circumventing the radiation pressure barrier via disk accretion}
\shortauthors{Kuiper et al.}

\begin{document}

\title{Circumventing the radiation pressure barrier in the formation of massive stars via disk accretion}


% \author{
% Rolf\ Kuiper\altaffilmark{1,2}, 
% Hubert\ Klahr\altaffilmark{2}, 
% Henrik\ Beuther\altaffilmark{2} 
% \and 
% Thomas\ Henning\altaffilmark{2}
% }
% \email{kuiper@mpia.de}
\author{Rolf\ Kuiper\altaffilmark{1,2}}
\email{kuiper@astro.uni-bonn.de}
\and 
\author{Hubert\ Klahr\altaffilmark{2}}
\and 
\author{Henrik\ Beuther\altaffilmark{2}}
\and 
\author{Thomas\ Henning\altaffilmark{2}}
\altaffiltext{1}
%{Argelander-Institut f\"ur Astronomie, Rheinische Friedrich-Wilhelms-Universit\"at Bonn, Auf dem H\"ugel 71, D-53121 Bonn, Germany} 
{Argelander Institute for Astronomy, Bonn University, Auf dem H\"ugel 71, D-53121 Bonn, Germany} 
\altaffiltext{2}
%{Max-Planck-Institut f\"ur Astronomie, K\"onigstuhl 17, D-69117 Heidelberg, Germany}
{Max Planck Institute for Astronomy, K\"onigstuhl 17, D-69117 Heidelberg, Germany}


%% generally not more than 250 words

\begin{abstract}
We present radiation hydrodynamics simulations of the collapse of massive pre-stellar cores.
We treat frequency dependent radiative feedback from stellar evolution and accretion luminosity at a numerical
resolution down to 1.27~AU.
In the 2D approximation of axially symmetric simulations,
it is possible for the first time 
to simulate the whole accretion phase 
(up to the end of the accretion disk epoch)
for the forming massive star
and 
to perform a broad scan of the parameter space.
Our simulation series show evidently the necessity 
to incorporate the dust sublimation front 
to preserve the high shielding property of massive accretion disks.
While confirming the upper mass limit of spherically symmetric accretion,
our disk accretion models show a persistent high anisotropy of the corresponding thermal radiation field.
This yields to the 
growth of the 
highest-mass stars ever formed in multi-dimensional radiation hydrodynamics simulations,
far beyond the upper mass limit of spherical accretion.
Non-axially symmetric effects are 
not necessary to sustain accretion.
The radiation pressure 
launches a stable bipolar outflow,
which grows in angle with time as 
presumed 
from observations.
For an initial mass of the pre-stellar host core of
60, 120, 240, and 480\Msol
the masses of the final stars formed in our simulations add up to
% TODO: update no.
28.2, 56.5, 92.6, and at least 137.2\Msol respectively.
\end{abstract}


\section{Introduction}
\label{sect:Introduction}
The understanding of massive stars still suffers from the lack of a generally accepted formation scenario.
Despite the strong limitations in observations of massive star forming regions compared to their low-mass counterpart,
past studies obtained common features of star formation,
suggesting that the formation of massive stars can in first order be treated analog to low-mass star formation.
All epochs of the classical picture of star formation,
such as
gravitationally collapsing massive cores 
\citep[e.g.][]{Ho:1986p4496, Keto:1987p4503, Zhang:1997p4579, Birkmann:2007p11062}
and
collimated jets as well as wide-angle bipolar outflows 
\citep[e.g.][]{Henning:2000p11099, Zhang:2002p8473, Zhang:2007p8480, Wu:2005p8491, Beuther:2005p1788}
were observed.
Even
circumstellar disks,
e.g.~IRAS~20126+4104 
\citep{Cesaroni:1997p4541, Zhang:1998p4666, Cesaroni:2005p4567} 
and 
AFGL~490 
\citep{Harvey:1979p11170, Torrelles:1986p11165, Chini:1991p11158, Davis:1998p11151, Lyder:1998p11169,
Schreyer:2002p11145, Schreyer:2006p11146},
or rather
large scale toroid
\citep{Beltran:2004p4711, Beltran:2005p4715}
and flattened rotating structures 
\citep{Beuther:2005p1793, Beuther:2008p4374}
could be revealed.
Reviews of observations related to a proposed picture of evolutionary sequences
of massive star formation are given in \citet{Beuther:2007p1908} or
\citet{Zinnecker:2007p752}.

Previous theoretical models focused on different points of view,
namely the competitive accretion 
\citep{Bonnell:1998p1416, Bonnell:2002p4306, Bonnell:2004p4756, Bate:2009p12138, Bate:2009p12130} 
and the turbulent core model 
\citep{McKee:2003p1221},
but they agree on the formation of accretion disks.

If the formation of high-mass stars is therefore treated as a scaled-up version of low-mass star formation,
a special feature of these high-mass proto-stars is the interaction of the accretion flow 
with 
the strong irradiation emitted by the newborn stars due to their short Kelvin-Helmholtz contraction timescale
\citep{Shu:1987p1616}. 
Previous one-dimensional studies 
\citep[e.g.][]{Larson:1971p1210, Kahn:1974p1200, Yorke:1977p1358, Wolfire:1987p539, Edgar:2004p15} 
agree on the fact that 
the growing radiation pressure potentially stops and reverts the accretion flow onto a massive star
yielding an upper mass limit of approximately 40\Msol.

But this radiative impact strongly depends on the geometry of the stellar environment
\citep{Nakano:1989p1267}.
The possibility was suggested to overcome this radiation pressure barrier
via the formation of a long-living massive circumstellar disk,
which forces the generation of a strong anisotropic feature of the thermal radiation field.
Earlier investigations by \citet{Yorke:2002p1} tried to identify such an anisotropy,
which they called the ``flashlight effect'',
in two-dimensional axially and midplane symmetric radiation hydrodynamics simulations, similar to our own.
Their simulations show an early end of the disk accretion phase shortly after its formation due to strong radiation
pressure.
The final star masses are only marginally higher than the mass limit in spherical symmetry,
even if the frequency dependence of the radiation is considered.
The radiative feedback was treated under the flux limited diffusion approximation
(hereafter called FLD), 
but computed for several frequency bins.
Due to the high computational cost of the frequency dependent FLD solver in
\citet{Yorke:2002p1},
it was unfortunately not possible to study the reason for the early fate of the disk accretion phase in detail.
\citet{Krumholz:2009p10975} 
stated that the circumstellar disk in the simulations of 
\citet{Yorke:2002p1} 
lost its shielding property because the disk region cannot be fed in axially symmetric configuration.
Contrary to the stable radiation pressure driven outflows in \citet{Yorke:2002p1},
they discovered in their own three-dimensional (frequency averaged) radiation hydrodynamics simulations
\citep{Krumholz:2007p1380, Krumholz:2009p10975} 
an instability in the outflow region,
leading to further accretion onto the disk.
They propose that this so-called 
``3D radiative Rayleigh-Taylor instability'' 
requires non-axially symmetric modes to occur
\citep{Krumholz:2009p10975}
and that radiation pressure therefore cannot halt a flow of gas and dust in any direction.
This explanation suffers from a lack of physical arguments 
and requires further detailed investigation 
due to the fact that
first the classical Rayleigh-Taylor instability is a two-dimensional instability
and 
secondly radiation cannot simply be treated as a fluid in general
(as it is in the sense of a Rayleigh-Taylor instability).
At least it remains unclear if this instability is the most important one.
With the help of the unstable polar region,
the most massive star in their simulations grows up to 41.5\Msol with an ongoing accretion phase,
as the simulation is not finished yet.

Contrary to this explanation of the short accretion phases in the two-dimensional simulations by \citet{Yorke:2002p1},
we demonstrate here in detail the need of including the radiation physics at the dust sublimation front of the forming
star to compute the correct anisotropy of the re-emitted radiation by dust grains.
Due to the huge sink cells used in the simulations by \citet{Yorke:2002p1},
the luminosity directly acts onto a disk region far beyond the actual dust sublimation front.
The interaction of the radiation with the accretion flow at the dust sublimation front is therefore artificially
shifted to the radius of the sink cell, 
where the circumstellar disk would originally be shielded from the direct stellar irradiation. 
In this region
the optical depth of the IR flux in the radial direction is much smaller 
than at the realistic location of the dust sublimation front,
yielding a much higher fraction of isotropy of the radiation field.
As shown in the well-established spherical symmetric studies,
this isotropic radiation field is able to stop and revert the accretion flow onto the forming star.

Our simulations focus on the accretion onto a single massive star in the center of the core.
We
incorporate the dust sublimation front of the forming star,
resolve the vicinity of the star down to 1.27~AU,
and evolve the system up to the order of $10^5$~yr 
(ten times longer than ever studied before),
including the whole disk accretion phase of the forming star.
We scan the broad parameter space 
of numerical configurations
as well as
different initial conditions
in several simulation series.

In the following Sect.~\ref{sect:physics}, 
we describe the details of the self-gravity radiation hydrodynamics code in
use. The physical initial conditions of the pre-stellar cores
as well as the numerical configuration of the simulations are described in Sect.~\ref{sect:InitialConditions}.
In Sects.~\ref{sect:1D} and \ref{sect:2D}, 
we present the results of one- and two-dimensional radiation hydrodynamics simulations 
of massive pre-stellar core collapses respectively, 
focusing on the radiative feedback onto the accretion flow while resolving the dust sublimation front. 
A discussion (Sect.~\ref{sect:discussion}) 
and summary of the most important results (Sect.~\ref{sect:conclusion})
including 
a comparison of our results to the simulations by
\citet{Yorke:2002p1} and \citet{Krumholz:2007p1380, Krumholz:2009p10975},
a discussion of our assumptions,
as well as a brief outlook on the future direction of this research project,
e.g.~3D simulations,
close this publication.

\section{Physics and numerics}
\label{sect:physics}
In this section, 
we outline the ingredients and default numerical configuration of the 
self-gravity radiation hydrodynamics code we use to model the collapse of massive pre-stellar cores. 
The first 
Sect.~\ref{sect:discretization} 
comprises the motivation for our choice of a grid in spherical coordinates
and highlights the step forward in resolution 
we obtain in our simulations compared to previous research.
The following two 
Sects.~\ref{sect:hydrodynamics} and 
\ref{sect:viscosity} describe the features and the configuration of the hydrodynamics solver
including full tensor viscosity, 
for which we use the open source magneto-hydrodynamics code Pluto3 \citep{Mignone:2007p3421}.
Further sections describe our newly developed modules for 
self-gravity (Sect.~\ref{sect:gravity}) 
and frequency dependent approximate radiation transport (Sect.~\ref{sect:radiation}).
We close this section with the description of 
the pre-calculated, tabulated dust (Sect.~\ref{sect:dust}) 
and stellar evolution model (Sect.~\ref{sect:star}) used in the simulations.

\subsection{Discretization of the computational domain}
\label{sect:discretization}
\begin{figure*}[p]
\begin{center}
\subfloat[Global image of the total computational domain up to the outer radius of $r_\mathrm{max} = 0.1$ pc.]{
\includegraphics[width=0.5\textwidth]{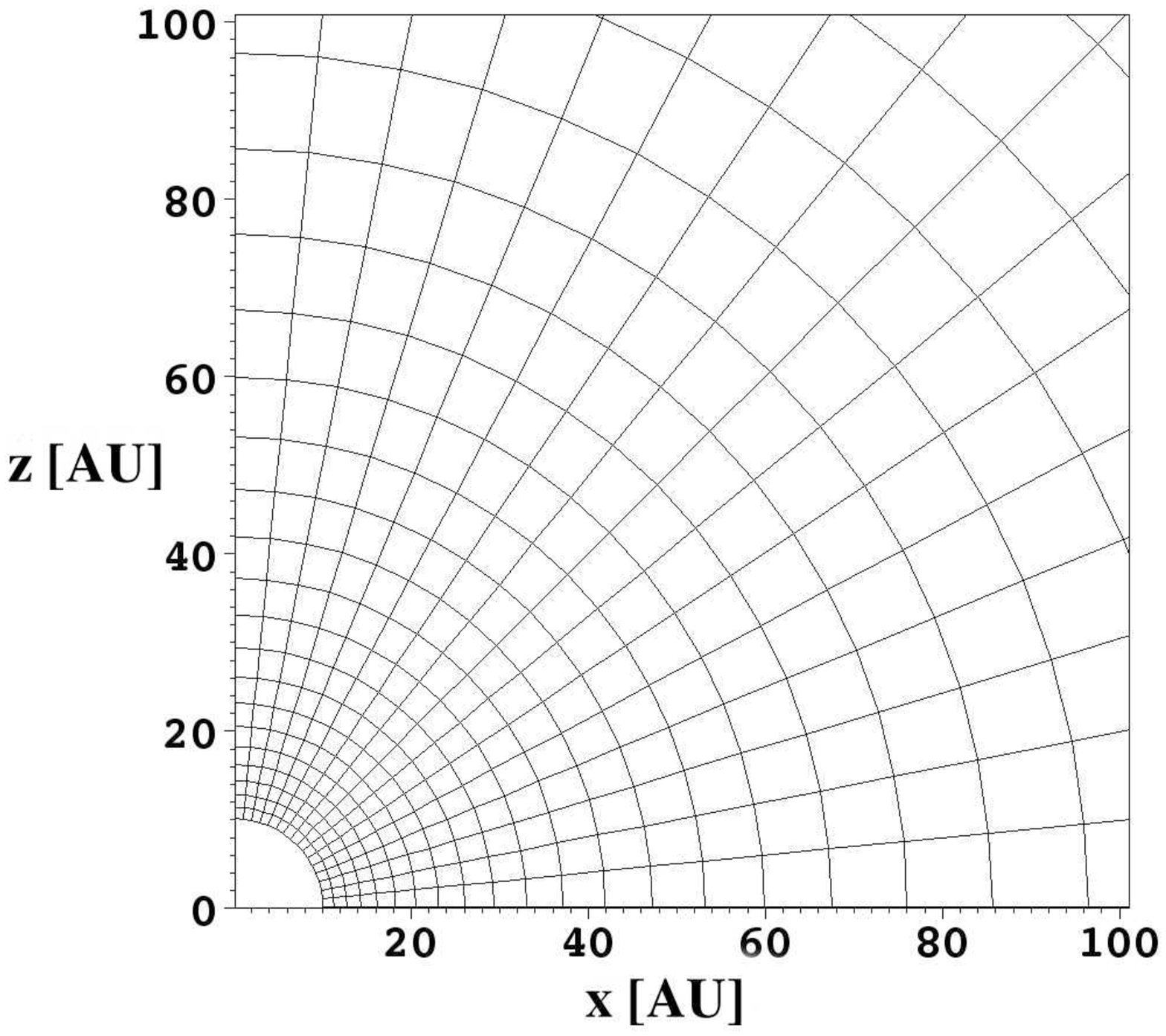}}
\hspace{1mm}
\subfloat[Zoom-in image of the central 100~AU~x~100~AU.
The innermost cells have a resolution of $\Delta r$~x~$r~\Delta \theta =$1.27~AU~x~1.04~AU.]{
\includegraphics[width=0.4565\textwidth]{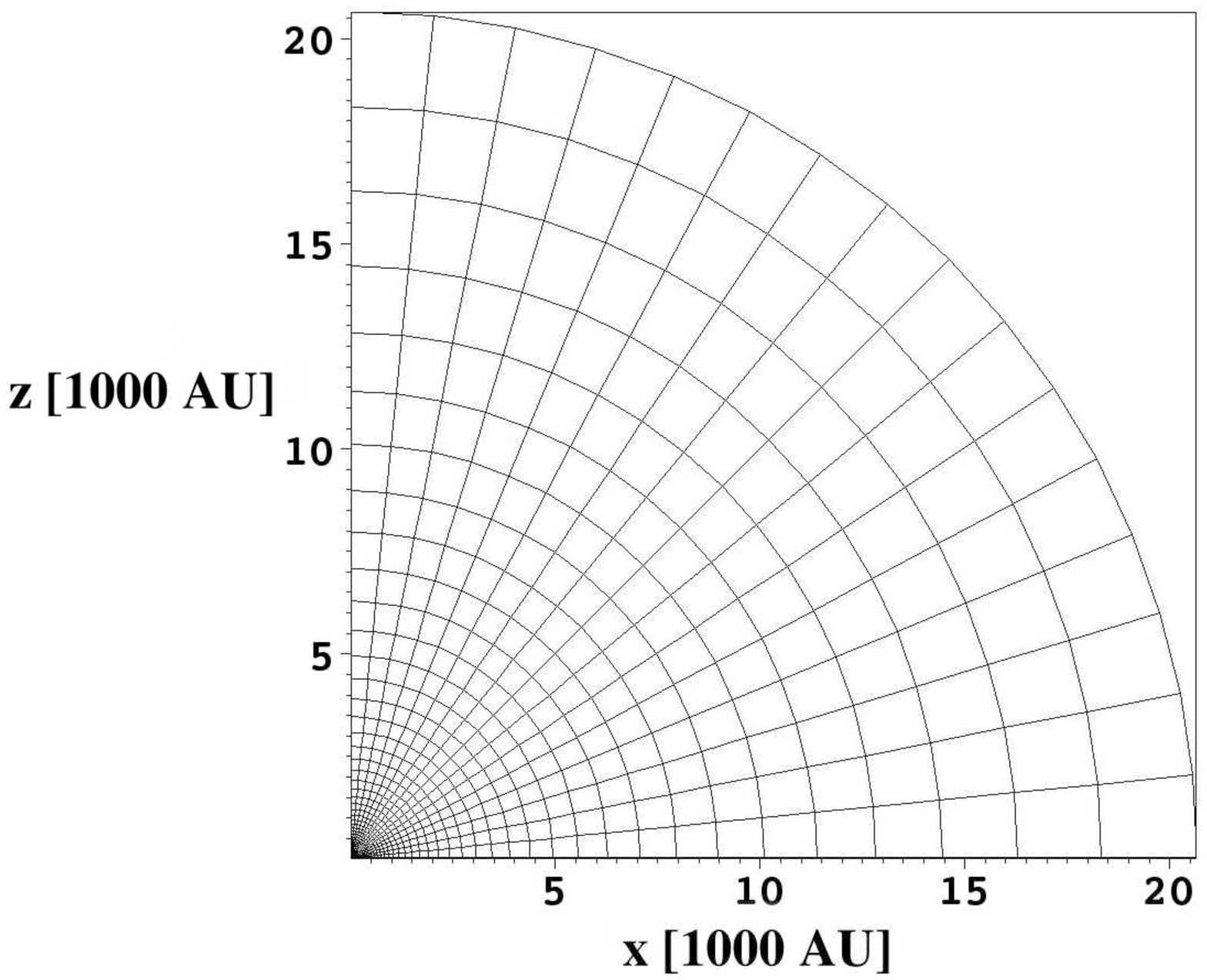}}
\caption{
Two-dimensional grid (64~x~16) in spherical coordinates with logarithmically increasing radial resolution, 
a central
sink cell of radius $r_\mathrm{min} = 10$~AU, 
and an outer boundary at $r_\mathrm{max} = 0.1$~pc. 
}
\label{Grid_low-res}
\end{center}
\end{figure*}

In our simulations, 
we use a time independent grid in spherical coordinates with logarithmically increasing radial resolution towards the
center. 
The usage of a radially increasing resolution towards the center guarantees the possibility 
to study the radiative feedback in the central core regions 
down to a minimum grid cell size of $\Delta r$~x~$r~\Delta \theta =$1.27~AU~x~1.04~AU.
An example of such a two-dimensional grid
is displayed in Fig.~\ref{Grid_low-res}. 
This type of grid is well adapted for the analysis of the
interaction of an accretion flow onto a massive star along with the stellar irradiation it generates, 
because the stellar gravity as well as the stellar radiative force are aligned with the radial coordinate axis. 
Furthermore,
in contrast to e.g.~cartesian coordinates, the usage of spherical coordinates guarantees a strict
angular momentum conservation.
The polar discretization $\Delta \theta$ of the 
grid is uniformly fixed 
and covers an angle of $\pi / 2$ from the top polar axis to the forming disk midplane, 
assuming midplane symmetry as in
\citet{Yorke:2002p1}. 
The polar 
resolution $r~\Delta \theta$ of the spherical grid automatically increases towards the regions of interest around
the centrally forming star, 
where high resolution is desired. 
To even enhance this
focus on the inner parts of the pre-stellar core and saving computational effort in the outer parts far away from the
dust radiation interaction layer,
we choose a logarithmically increasing radial resolution of the grid.
Thus the non-adaptive grid setup impeded the study of potential fragmentation in the outer core regions.
The radial resolution
$\Delta r$ at a radius $r$ of a computational domain with $N_r$ grid cells in the radial direction is given by
\begin{equation}
\Delta r \left(r\right) = r ~ \left(10^f - 1 \right)
\end{equation}
with 
$f = \log\left(r_\mathrm{max}/r_\mathrm{min}\right) / N_r$, 
where $r_\mathrm{min}$ and $r_\mathrm{max}$ represent the inner and outer radius of the
computational domain. 
A comparison of our achieved resolution to previous massive pre-stellar core collapse simulations by
\citet{Yorke:2002p1}
and
\citet{Krumholz:2007p1380, Krumholz:2009p10975}
is given in Table~\ref{resolution}.
\begin{deluxetable}{l c c c}
\tablecaption{Resolution of different radiation hydrodynamics simulations of a
collapse of a slowly rotating massive pre-stellar core
\label{resolution}}
\tablehead{
&
\multicolumn{2}{c}{Resolution in AU in regions of} & 
\colhead{Radius of} \\
\colhead{Authors} &
\colhead{lowest resolution}  &
\colhead{highest resolution} & 
\colhead{sink cells (AU)}
}
\startdata 
\citet{Yorke:2002p1}       &  $320^2$           & $80^2$ & 80\\
\citet{Krumholz:2007p1380} &  $966^3$           & $7.5^3$ & 0 - 30\\
\citet{Krumholz:2009p10975} &  $645^3$           & $10^3$  & 0 - 40\\
This study~1D           & 1540               &  0.08 & 1.0  \\
This study~2D           & 2319 x 1911        &  1.27 x 1.04 & 10.0\\
\enddata 
\tablecomments{
The simulations of 
\protect\citet{Yorke:2002p1} 
were performed on a non-adaptive two-dimensional grid in cylindrical coordinates with three levels of refinement. 
The given resolution 
$\left(\Delta r \mbox{ x } \Delta z\right)$ 
of 
\protect\citet{Yorke:2002p1} 
represents the case of a
$M_\mathrm{core} = 60 \mbox{ M}_\odot$ 
pre-stellar core.
The resolution for the lower mass 
$M_\mathrm{core} = 30 \mbox{ M}_\odot$ 
collapse was a factor of two better.
The resolution for the higher mass 
$M_\mathrm{core} = 120 \mbox{ M}_\odot$ 
collapse was a factor of two worse.
The simulations of 
\protect\citet{Krumholz:2007p1380, Krumholz:2009p10975} 
were performed on a
three-dimensional cartesian adaptive mesh refinement (AMR) grid.
The given resolution 
$\left( \Delta x \mbox{ x } \Delta y \mbox{ x } \Delta z \right)$ 
represents the lowest and highest resolution during the simulation. 
The radiation from the sink cells is added to their computational grid using a smooth weighting function inside the
so-called accretion radius, 
which is four times the highest resolution of the grid.
\citep{Krumholz:2005p859}.
The resolution of our own grids in spherical coordinates is given in units of arc length
$\left( \Delta r \mbox{ x } (r~\Delta \theta) \right)$.
}
\end{deluxetable} 

The forming high-mass proto-stellar object at the center of the core is represented by a dedicated stellar evolution
model (presented in Sect.~\ref{sect:star}) inside the central sink cell with radius $r_\mathrm{min}$ 
at the origin of the coordinate system using pre-calculated stellar
evolutionary tracks for accreting high-mass proto-stars \citep{Hosokawa:2009p12591}.

\subsection{Hydrodynamics}
\label{sect:hydrodynamics}
To follow the motion of the gas, 
we solve the equations of compressible hydrodynamics
\begin{eqnarray}
\label{eq:hydrodynamics_density}
\partial_t ~ \rho + \vec{\nabla} \cdot \left(\rho ~ \vec{u}\right) &=& 0 \\
\label{eq:hydrodynamics_momentum}
\partial_t \left(\rho ~ \vec{u}\right) + \vec{\nabla} \left(\rho ~ \vec{u} ~ \vec{u} + P\right) 
&=& \rho ~ \vec{a}
\\
\label{eq:hydrodynamics_energy}
\partial_t ~ E + \vec{\nabla} \cdot \left(\left(E + P\right) ~ \vec{u}\right) 
&=& \rho ~ \vec{u} \cdot \vec{a} - \vec{\nabla} \vec{F}_\mathrm{tot}
\end{eqnarray}
with the acceleration source term 
$\vec{a} = \sum_i \vec{a}_i$, 
which includes the additionally considered physics to the equations of gas dynamics 
(Euler equations) 
such as shear viscosity $(\vec{a}_1)$, 
central gravity of the forming star $(\vec{a}_2)$, 
self-gravity $(\vec{a}_3)$, 
and radiation transport and stellar radiative feedback $(\vec{a}_4)$. 
$\vec{F}_\mathrm{tot}$ denotes the flux of the total radiation energy density.
These additional components are described in the following subsections.
The evolution of 
the gas density $\rho$, 
velocity $\vec{u}$, 
pressure $P$, 
and total energy density $E$ 
is computed using the open source magneto-hydrodynamics code Pluto3 
\citep{Mignone:2007p3421}.

Pluto is a high-order Godunov solver code, 
i.e.~it uses a shock capturing Riemann solver within a conservative finite volume scheme. 
The numerical configuration of our simulations makes use of 
a Strang operator splitting scheme for the different dimensions 
\citep{Strang:1968p6618}.
Our default configuration consists further of a 
Harten-Lax-Van~Leer (hll) Riemann solver 
and a so-called `minmod' flux limiter 
using piecewise linear interpolation (plm) 
and a Runge-Kutta 2 (RK2) time integration, 
also known as the predictor-corrector-method, 
compare
\citet{vanLeer:1979p5193}. 
Therefore the total difference scheme is accurate to $2^\mathrm{nd}$ order in time and space.

To close the system of Eqs.~\eqref{eq:hydrodynamics_density} to \eqref{eq:hydrodynamics_energy},
we use an ideal gas equation of state
\begin{equation}
P = \left(\gamma - 1\right) E_\mathrm{int},
\end{equation}
which relates the gas pressure $P$ to the internal energy 
$E_\mathrm{int} = E - 0.5 \rho u^2$.
The adiabatic index $\gamma$ is set to $5/3$.

To limit the range of densities, 
the so-called floor value of the density is chosen to be $\rho_0 = 10^{-21} \rhocgs$.
This floor value occurs during the simulations only in regions 
where the radiation pressure driven outflow is depleting the density of the corresponding grid cells in the
radially outward direction. 
Thus, 
the choice of the floor value does not influence the level of accretion onto the newly forming star we 
investigate.

The various sources of additional acceleration 
$\vec{a} = \sum_i \vec{a}_i$ 
that enter the equations of hydrodynamics in Eqs.~\eqref{eq:hydrodynamics_momentum} and
\eqref{eq:hydrodynamics_energy} are discussed in the following sections and include 
viscosity, 
gravity of the central star as well as self-gravity of the core, 
and radiative feedback.

\subsection{Viscosity}
\label{sect:viscosity}
In the two-dimensional simulations we consider physical shear viscosity of the circumstellar disk medium to mimic the effect of angular momentum transport (via e.g.~the magneto-rotational instability, spiral arms, disk winds and jets). 
Two-dimensional axially symmetric simulations without any shear viscosity yield the formation of ring instabilities in the circumstellar disk. 
The rings would be unstable if non-axially symmetric modes were allowed, leading to the formation of spiral arms and therefore to angular momentum transport as discussed by \citet{Yorke:1995p1426}.

Full physical tensor viscosity is included in the current version of the open source magneto-hydrodynamics code Pluto3 \citep{Mignone:2007p3421}.
Viscosity enters the equations of hydrodynamics in Eqs.~\eqref{eq:hydrodynamics_momentum} and \eqref{eq:hydrodynamics_energy} as an additional source of acceleration
\begin{equation}
\vec{a}_1 = \vec{\nabla} \Pi.
\end{equation}
The components of the viscous stress tensor $\Pi$ are given (in cartesian coordinates) by
\begin{equation}
\Pi_{ij} = \eta \left(\partial_j u_i + \partial_i u_j - \frac{2}{3} \delta_{ij} \partial_k u_k \right) 
+ \eta_\mathrm{b} \delta_{ij} \partial_k u_k
\end{equation}
with the shear viscosity $\eta$, the bulk viscosity $\eta_\mathrm{b}$, and the Kronecker symbol $\delta_{ij}$.
Further details on the analytical treatment of viscosity can e.g.~be found in \citet{DLandau:1987p5480}.
We assume for the bulk viscosity $\eta_\mathrm{b} = 0$.
\newline
The (shear) viscosity 
\begin{equation}
\eta = \rho ~ \nu
\end{equation}
is described via the so-called $\alpha$-parameterization of \citet{Shakura:1973p3060}, in which the dynamical viscosity $\nu$ is set proportional to the product of a typical velocity and length scale of the system under investigation, here the local sound speed $c_\mathrm{s}$ and pressure scale height $H$:
\begin{equation}
\label{eq:shakurasunyaev}
\nu = \alpha ~ c_\mathrm{s} ~ H
\end{equation}
We further approximate the local pressure scale height $H$ by
\begin{equation}
\label{eq:pressurescaleheight}
H = \frac{c_\mathrm{s}}{\Omega_\mathrm{K}(r)}
\end{equation}
with the keplerian angular velocity
\begin{equation}
\Omega_\mathrm{K}(r) = \sqrt{\frac{G M(r)}{r^3}}
\end{equation}
derived from the equilibrium between gravity and the centrifugal force.
The mass $M(r)$ inside the radius $r$ is calculated by the spatial integral of the density distribution plus the central stellar mass $M_*$ inside the sink cell:
\begin{equation}
\label{eq:includedmass}
M(r) = M_* + 2\pi ~ \int_0^r dr \int_0^\pi d\theta ~ \rho(r,\theta) ~ r^2 \sin(\theta).
\end{equation}
Using the relation \eqref{eq:pressurescaleheight} we substitute the local sound speed in Eq.~\eqref{eq:shakurasunyaev}
yielding
\begin{equation}
\nu = \alpha ~ \Omega_\mathrm{K}(r) ~ H^2.
\end{equation}
Introducing the dimensionless parameter $H/R$, the aspect ratio of the circumstellar disk, leads to
\begin{equation}
\nu = \alpha ~ \Omega_\mathrm{K}(r) ~ R^2 \left(\frac{H}{R}\right)^2
\end{equation}
with the cylindrical radius $R = r ~ \sin(\theta)$.
\vONE{
If the viscosity is an effect of turbulent transport of angular momentum, e.g.~by the magneto-rotational instability \citet{Balbus:1991p3799, Hawley:1991p11900, Balbus:2003p11906} or the baroclynic instability \citet{Klahr:2003p2794}, it is observed that the strength of the stresses is proportional to the thermal pressure. 
This is the fundamental assumption of the $\alpha$-Ansatz by \citet{Shakura:1973p3060}.
This relation (Eq.~\eqref{eq:shakurasunyaev}) holds because hotter and thicker disks can support higher levels of turbulence.
The situation is reversed for self-gravitating disks.
Here, hot disks are usually Toomre stable and will not produce gravito-turbulence. 
On the contrary, the disks will cool down to the marginally unstable Toomre values and establish a turbulent state where the level of turbulence is set by the equilibrium of energy release and radiative cooling \citep{Gammie:2001p14271}.
For that reason, we choose a viscosity prescription independent on the actual disk temperature (e.g.~a fixed $H/R$ ratio of $0.1$ and $\alpha = 0.3$) but only on the local mean (keplerian) rotation profile.
This way, we ensure that cool and thin disks can obtain the high viscosity values they deserve.
Our Ansatz is equivalent to the so-called $\beta$-viscosity Ansatz for self-gravitating disks by \citet{Duschl:2000p14177}, which is also independent on temperature, with the $\beta$-parameter of $\beta \approx 3*10^{-3}$.
}
%ONEBoth unit-free parameters $H / R = 0.1$ and $\alpha = 0.3$ are constant in time and space for the majority of our simulation runs.

\subsection{Gravity}
\label{sect:gravity}
The calculation of the gravitational potential $\Phi$ is split into 
the gravity of the central star in the sink cell
$\Phi_*$ 
and the self-gravity of the mass in the computational domain 
$\Phi_\mathrm{sg}$:
\begin{equation}
\Phi = {\Phi_{*}} + \Phi_\mathrm{sg}
\end{equation}
The associated accelerations $\vec{a}_2+\vec{a}_3$ enter the hydrodynamics as a source term for momentum and energy in Eqs.~\eqref{eq:hydrodynamics_momentum} and \eqref{eq:hydrodynamics_energy}. 

The acceleration vector 
$\vec{a}_2$ 
of the gravity of the central star
%ONEwhich enters the conservation laws of hydrodynamics in Eqs.~\eqref{eq:hydrodynamics_momentum} and \eqref{eq:hydrodynamics_energy}, 
is given analytically by
\begin{equation}
\vec{a}_2 = - \vec{\nabla} ~ {\Phi_{*}} = \vec{\nabla} ~ \frac{G ~ M_*}{r} = \partial_r ~ \frac{G ~ M_*}{r} \vec{e}_r =
- \frac{G ~ M_*}{r^2} \vec{e}_r.
\end{equation}
Such external gravity 
(from point sources) 
is supported in Pluto3 by defining the gravitational potential 
$\Phi_*$ 
or the resulting acceleration vector 
$\vec{a}_2$. 
\newline
The acceleration 
$\vec{a}_3$ 
due to self-gravity is given by
\begin{equation}
\vec{a}_3 = - \vec{\nabla} ~ \Phi_\mathrm{sg},
\end{equation}
in which the gravitational potential 
$\Phi_\mathrm{sg}$ 
is determined via Poisson's equation:
\begin{equation}
\label{eq:Poisson}
\vec{\nabla}^2 \Phi_\mathrm{sg} = 4 \pi ~ G ~ \rho.
\end{equation}
We implemented a solver of Poisson's equation into our version of the Pluto code in a modular fashion. 
The module solves the equation via a diffusion ansatz.
The desired approximate matrix inversion is done using the so-called GMRES method.
%ONEwhich is also used for the FLD Eq.~\eqref{eq:rad_diffusion}.
The accuracy of the Poisson solver,
i.e.~the abort criterion for the approximate matrix inversion,
is choosen to 0.001\% relative accuracy of the gravitational potential 
$\left(\Delta \Phi_\mathrm{sg} / \Phi_\mathrm{sg} \le 10^{-5}\right)$.

The outer radial boundary values of the gravitational potential are calculated via a Taylor expansion of the density
distribution, 
described for example in
\citet{Black:1975p3742}.
Several tests we performed indicate that it is sufficient to just account for the monopole solution of the Taylor
expansion, 
i.e.~the total mass of the core. 
In our default configuration of the pre-stellar cores (see Sect.~\ref{sect:InitialConditions}),
the mass distribution is perfectly spherically symmetric at the beginning of the simulation and afterwards becomes
highly concentrated in the inner region of the computational domain far away from the outer boundary, 
both yield analytically the monopole solution at the outer boundary. 
To control the resolution, 
which is necessary to resolve the physics of self-gravity correctly, 
e.g.~preventing artificial fragmentation,
we monitor the so-called Truelove criterion,
derived in 
\citet{Truelove:1997p10742}.
The criterion requires to resolve the Jeans length
\begin{equation}
\lambda_\mathrm{J} = \sqrt{\frac{\pi ~ c_\mathrm{s}^2}{G ~ \rho}}
\end{equation}
at least by a priorily defined number of grid cells. 
The inverse of the number of necessary grid cells per Jeans length is the so-called Jeans number
\begin{equation}
J = \frac{\Delta x}{\lambda_\mathrm{J}}.
\end{equation}
\citet{Truelove:1997p10742} 
suggested at least a Jeans number of $J \le 0.25$.
The worst case during our simulations occurs in the high-density region 
around the forming star with approximately a maximum density of
$\rho \le 10^{-11} \rhocgs$,
a minimum temperature of
$T \approx 100 \mbox{ K}$
and a resolution of the order of 
$\Delta r \approx 1 \mbox{ AU}$.
This leads to a Jeans number of
$J \approx 0.09$, 
i.e.~the Jeans length $\lambda_\mathrm{J}$ is at least resolved by 11 grid cells.

%ONEThis coupling to the hydrodynamics was cross-checked in dynamical collapse tests. 

\subsection{Radiation transport}
\label{sect:radiation}
The importance of the frequency dependence of the stellar spectrum when calculating the radiative feedback of a massive
star was already shown in radiation hydrodynamics studies by 
\citet{Yorke:2002p1} 
and \citet{Edgar:2003p6} 
as well as in radiation transport simulations without hydrodynamics by
\citet{Krumholz:2005p867}. 
On the other hand, 
no frequency dependent radiation hydrodynamics study related to massive star formation was carried out since the work by 
\citet{Yorke:2002p1} 
in more than one dimension
and
due to the huge computational overhead of their frequency dependent FLD routine it was
neither possible to study a large number of different initial conditions 
(to scan the parameter space), 
nor to perform high-resolution simulations of the accretion process.

Furthermore, stellar radiative feedback onto the dynamics of the environment plays a crucial role in the formation of massive stars.
On the one hand, the heating will probably prevent further fragmentation of the cloud by enhancing the Jeans mass \citep[e.g.][]{Krumholz:2007p1380}.
On the other hand, the dusty environment feels the radiative force when absorbing the radiation due to momentum conservation, which potentially stops the accretion process for highly luminous massive stars. 
%ONETherefore, it is necessary to compute the correct radiative flux (and its derivative, the radiative force) in addition to the temperature distribution in such simulations.

To study the radiative feedback of massive stars on their own accretion stream in one-, two-, and
three-dimensional simulations we implemented a 
fast, 
robust, 
and accurate frequency dependent radiation transport solver
in spherical coordinates into our version of the Pluto code.
To achieve a fast solver for the frequency dependent problem we split the radiation field into the stellar irradiation
and thermal dust emission.
%ONEThe robustness of the solver was realized by using a modern Krylov subspace solver for sparse linear matrix equations provided by the scientific open source library called `PETSc' \citep{petsc-web-page}.
%ONEThe accuracy of the final radiation transport module was tested in detail against the standard radiation benchmark test by \citet{Pascucci:2004p39} for Monte-Carlo or ray-tracing radiative transfer solvers in \citet{Kuiper:2010p12874}.
%ONEWe found that it is necessary to consider the frequency dependence of the stellar irradiation feedback to reconstruct a reasonable approximation to the spatial temperature distribution.
%ONEThe approximation results in a large reduction in computing time compared to Monte-Carlo based radiative transfer.
\vONE{
The basic methodology of the hybrid scheme is to perform a frequency dependent ray-tracing step for the stellar irradiation and shift the re-emission of the photons by the dusty environment to a frequency averaged flux limited diffusion solver in the equilibrium temperature approximation.
A derivation of the hybrid scheme and numerical details of the implementation of this newly developed radiation transport method are given in \citet{Kuiper:2010p12874},
including a detailed comparison of the method with the standard radiation transport benchmark test by \citet{Pascucci:2004p39}.
}

\subsection{Dust model}
\label{sect:dust}
\begin{figure}[htbp]
\begin{center}
\includegraphics[angle=0,width=\FigureWidth]{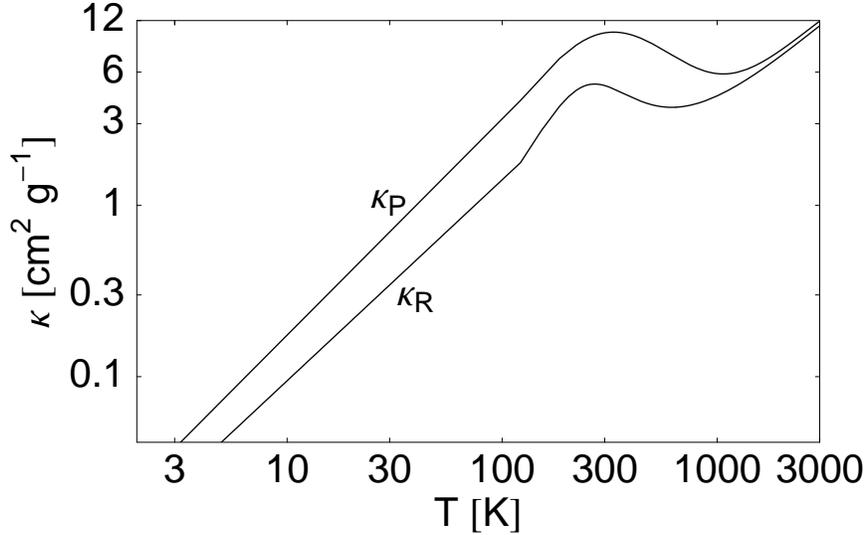}
\caption{
Calculated Rosseland $\kappa_\mathrm{R}$ and Planck $\kappa_\mathrm{P}$ mean opacities as a function of dust
temperature. 
Here, the possible evaporation of dust grains at high temperatures (and/or low densities) is neglected, 
but considered in the dust to gas mass ratio $M_\mathrm{dust}/M_\mathrm{gas}$ of each grid cell, 
cp.~Fig.~\protect\ref{Dust2GasRatio}.
}
\label{Opacities_vs_T}
\end{center}
\end{figure}

For the implementation of realistic mass absorption coefficients $\kappa\left(\nu\right)$ for the frequency dependent radiation transport module, 
we use an opacity table of \citet{Laor:1993p736}, including 79 frequency bins,
\vONE{
shown in Fig.~\ref{Opacities}.
\begin{figure}[htbp]
\begin{center}
\includegraphics[angle=270,width=\FigureWidth]{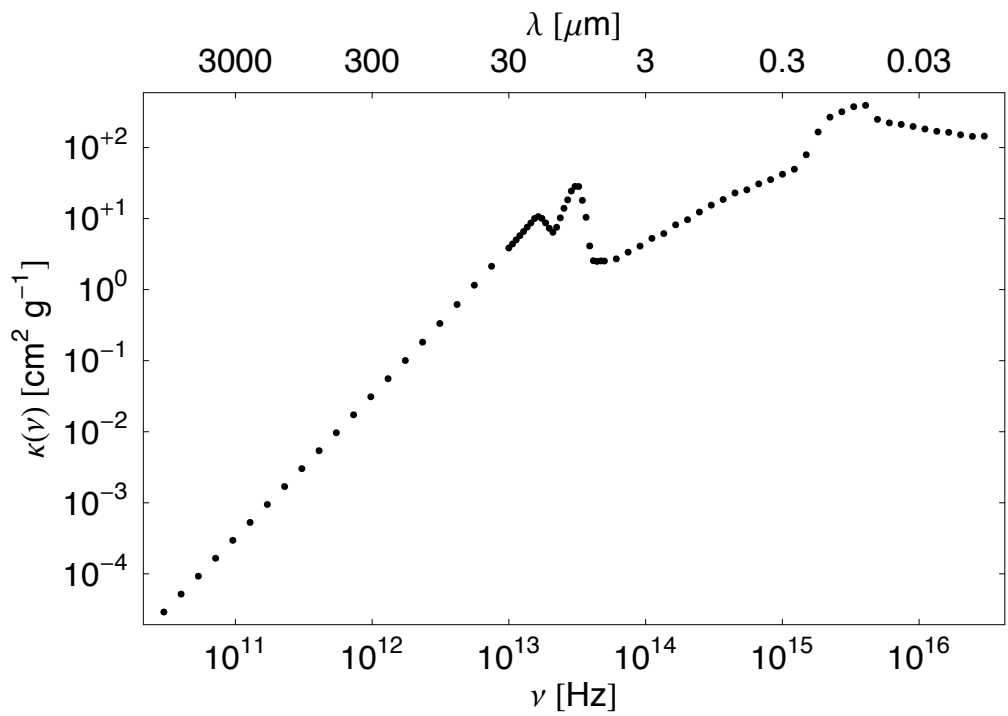}
\caption{
Frequency dependent mass absorption coefficients $\kappa(\nu)$ in tabulated form from 
\protect\citet{Laor:1993p736}.
}
\label{Opacities}
\end{center}
\end{figure}
%ONEEach dot in Fig.~\ref{Opacities} represents the mid-frequency of the corresponding frequency bin.
}
The opacity table covers the full frequency range from microwave and infrared radiation up to soft x-rays.
It describes a mixture of dust grains in the size range between $0.005 \mbox{ to } 10.0 \mbox{ } \mu \mbox{m}$.
The grains are taken to be spherical and consist out of amorphous silicate with a composition like that of olivine.
\vONE{
The model does not include ice mantles, which will roughly double the opacity at temperatures below $\sim100$~K \citep[e.g.][]{Ossenkopf:1994p4072}.
}
As shown in Fig.~\ref{Opacities} this dust grain mixture takes into account the strong
absorption/emission features at $9.7 ~ \mu\mbox{m}$ 
and $18 ~ \mu\mbox{m}$ observed in the interstellar medium.
The corresponding frequency averaged Planck- and Rosseland mean opacities are shown in Fig.~\ref{Opacities_vs_T} 
as a function of temperature. 

\begin{figure}[htbp]
\begin{center}
\includegraphics[angle=270,width=\FigureWidth]{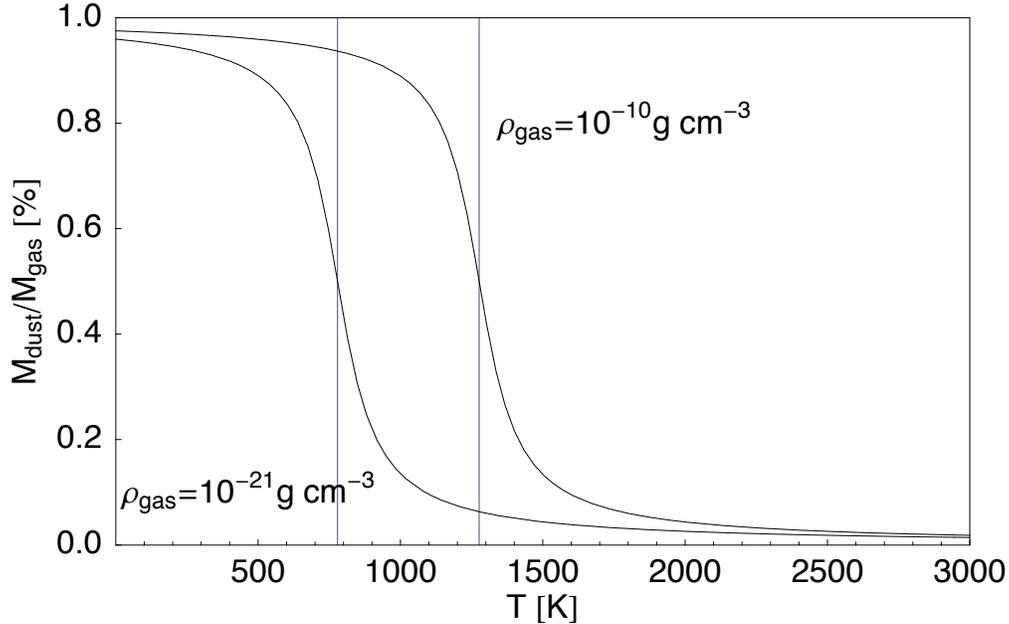}
\caption{
Transition slope of the local dust to gas mass ratio as a function of the temperature due to evaporation of dust grains
for two different gas densities. 
The vertical lines mark the corresponding evaporation temperatures.
}
\label{Dust2GasRatio}
\end{center}
\end{figure}

Aside from the dust opacities, 
the opacity of a given grid cell depends also linearly on the local dust to gas mass ratio. 
The initial dust to gas mass ratio $\left(M_\mathrm{dust}/M_\mathrm{gas}\right)_0 $ is fixed to 1\%.
Gas and dust is treated as a single fluid, 
so the dust to gas mass ratio only shrinks due to possible evaporation of the dust grains in hot regions 
(around the central massive star). 
The local evaporation temperature of the dust grains is calculated by using the formula of
\citet{Isella:2005p3014}
\begin{equation}
T_\mathrm{evap} = g \mbox{ } \rho^{~\beta}
\end{equation}
with 
$g = 2000 \mbox{ K}$, 
$\beta = 0.0195$, 
and the gas density $\rho$ given in $\mbox{g cm}^{-3}$.
The formula describes a power-law approximation to the evaporation temperatures $T_\mathrm{evap}$ determined by
\citet{Pollack:1994p3016}.
A smooth spatial and temporal transition of the associated dust to gas mass ratio between completely evaporated
and condensated regions is achieved via the transition function
\begin{equation}
\frac{M_\mathrm{dust}}{M_\mathrm{gas}}(\vec x) =
\left(\frac{M_\mathrm{dust}}{M_\mathrm{gas}}\right)_0
\left(
0.5 - \frac{1}{\pi}
\arctan \left( \frac{T(\vec x) - T_\mathrm{evap}(\vec x)}{100} \right)
\right).
\end{equation}
The transition slope is displayed in 
Fig.~\ref{Dust2GasRatio} 
as a function of the temperature for a high gas density of
$\rho = 10^{-10} \mbox{ g} \mbox{ cm}^{-3}$ 
as well as for the floor value of the density 
$\rho_0 = 10^{-21} \mbox{ g} \mbox{ cm}^{-3}$.

\subsection{Stellar evolution model}
\label{sect:star}
\begin{figure}[htbp]
\begin{center}
\includegraphics[angle=270,width=\FigureWidth]{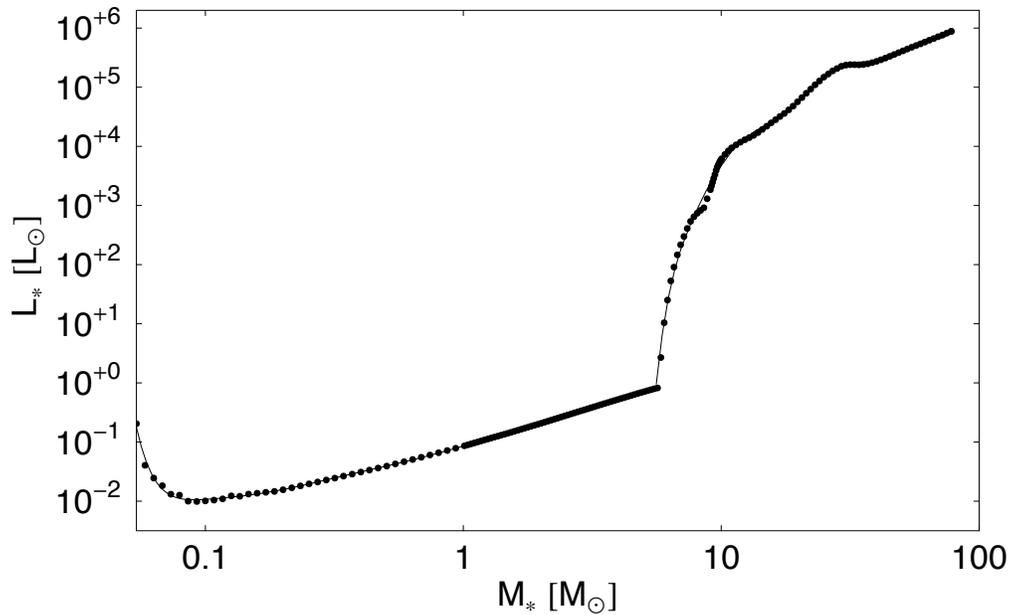}
\caption{
Polynomial fits to the stellar luminosity as a function of the stellar mass as calculated by
\protect\citet{Hosokawa:2009p12591}.
The data points represent an evolving massive star with an accretion rate of $10^{-3} \mbox{ M}_\odot \mbox{ yr}^{-1}$.
The mass range was split into two regimes above and below $5.5 \mbox{ M}_\odot$ (at the sharp bend) and each part is
fitted by a polynomial up to $10^\mathrm{th}$ order (solid lines). 
}
\label{Hosokawa}
\end{center}
\end{figure}

The evolution of the central star, 
described by a central sink cell, 
is coupled to the hydrodynamics of the
pre-stellar core by measuring the mass flux into the sink cell. 
The initial mass of the star at the beginning of the
simulation is simply given by the integral over the initial density distribution up to the radius $r_\mathrm{min}$ of the
sink cell and is therefore in all cases less than a few percent of $1 \mbox{ M}_\odot$. The mass, which enters the sink
cell during the hydrodynamics is assumed to be accreted onto the central star. 
From the mass flux $\rho \vec{u}$
into the sink cell during the timestep $\Delta t$ 
we calculate the accretion rate $\dot{M}$ onto the central star via
\begin{equation}
\dot{M} = 2 \pi \int_0^\pi d\theta \mbox{ } \rho \vec{u} \cdot \vec{e}_r ~ r_\mathrm{min}^2 \sin{\theta}. 
\end{equation}
Integrating the accretion rate $\dot{M}$ over the time yields the growth of the stellar mass $M_*$:
\begin{equation}
M_*(t+\Delta t) = M_*(t) + \int_t^{t+\Delta t} \dot{M} dt = M_*(t) + \dot{M} \Delta t.
\end{equation}
A potential decrease of the accretion rate due to outflows or jets 
(driven by magnetic forces) 
in the inner region below the sink cell radius, 
is currently not considered.
The total luminosity $L_\mathrm{tot}$ is given by the sum of the accretion luminosity $L_\mathrm{acc}$ and
the stellar luminosity $L_*$:
\begin{equation}
L_\mathrm{tot}(t) = L_\mathrm{acc}(t) + L_*(t).
\end{equation}
The accretion luminosity is directly calculated from the hydrodynamics simulation via
\begin{equation}
L_\mathrm{acc} =\frac{G M_*}{R_*} \dot{M}
\end{equation}
with the stellar radius $R_*$.
The stellar luminosity and the stellar radius are obtained via fits to the pre-calculated evolutionary tracks by
\citet{Hosokawa:2009p12591}.
These evolutionary tracks of massive stars depend on the stellar mass as well as on the actual accretion rate.
We use polynomial fits to the mass relation up to $10^\mathrm{th}$ order for separated mass ranges (an example of these
fits is shown in Fig.~\ref{Hosokawa}) and linear regression for the dependency on the \vONE{instantaneous} accretion rate. 
For stellar masses below the accessible data ($0.05 \mbox{ M}_\odot$ in the worst case) the stellar luminosity is
assumed to be negligible and the stellar radius is assumed to be constant up to the first data point.

Given the stellar radius and total luminosity, 
the stellar effective temperature $T_*$ is calculated from
\begin{equation}
L_\mathrm{tot} = 4 \pi ~ R_*^2 ~ \sigma_\mathrm{SB} ~ T_*^4.
\end{equation}

\section{Initial conditions and numerical configuration}
\label{sect:InitialConditions}
Using the newly developed modules of the self-gravity radiation hydrodynamics
code presented in Sect.~\ref{sect:physics}, we performed multiple simulations of collapsing massive pre-stellar cores.
Most of the simulations were performed either 
to scan the huge numerical parameter space of the setup to guarantee significant results 
or to explore individual physical initial conditions. 
An overview of the 22 simulations evaluated is presented in 
Table~\ref*{tab:1Druns} 
and
Table~\ref*{tab:2Druns} 
for one- and two-dimensional simulations respectively.

\begin{deluxetable}{l c c c c c c c}
\tablecaption{Overview of spherically symmetric massive pre-stellar core collapse simulations
\label{tab:1Druns}}
\tablehead{
&
&
& 
\colhead{Resol.} &
\colhead{$r_\mathrm{min}$} &
\colhead{$M_\mathrm{core}$} &
\colhead{$t_\mathrm{ff}$} &
\colhead{$t_\mathrm{end}$} \\
\colhead{Label} &
\colhead{Dim.}  &
\colhead{Grid cells} & 
\colhead{(AU)} &
\colhead{(AU)} &
\colhead{$(\mbox{M}_\odot)$} &
\colhead{(kyr)} &
\colhead{(kyr)}
}
\startdata 
\multicolumn{2}{l}{1D Convergence runs} & \multicolumn{6}{l}{Sect.~\ref{sect:1DConvergence}} \\
\hline
1D-Convergence32     & 1D &  32       & 0.36 &   1     &  60 & 67.6 & $310^{**}$
\\ 
1D-Convergence64     & 1D &  64       & 0.17 &   1     &  60 & 67.6 & $163^{**}$
\\ 
1D-Convergence128    & \multicolumn{7}{l}{see `1D-Mcore60Msol'} \\
1D-Convergence256    & 1D & 256       & 0.04 &   1     &  60 & 67.6 & $188^*$ \\
\multicolumn{2}{l}{1D rmin runs} & \multicolumn{6}{l}{Sect.~\ref{sect:1Drmin}} \\
\hline
1D-rmin1AU      & 1D & $99+128$  & 1.0    &  1      &  60 & 67.6 & $204^*$\\
1D-rmin5AU      & 1D & $95+128$  & 1.0    &  5      &  60 & 67.6 & $282^{**}$\\
1D-rmin10AU     & 1D & $90+128$  & 1.0    & 10      &  60 & 67.6 & $283^{**}$\\
1D-rmin80AU     & 1D & $20+128$  & 1.0    & 80      &  60 & 67.6 & $293^*$\\
\multicolumn{2}{l}{1D Mcore runs} & \multicolumn{6}{l}{Sect.~\ref{sect:1DMcore}} \\
\hline
1D-Mcore60Msol    & 1D & 128       & 0.08 &  1      &  60 & 67.6 & $218^{**}$\\
1D-Mcore120Msol   & 1D & 128       & 0.08 &  1      & 120 & 47.8 & $54^{**}$\\
1D-Mcore240Msol   & 1D & 128       & 0.08 &  1      & 240 & 33.8 & $39^{**}$\\
1D-Mcore480Msol   & 1D & 128       & 0.08 &  1      & 480 & 23.9 & $10^{**}$\\
\enddata 
\tablecomments{
The table is structured in blocks of topics and their corresponding sections.
For each run 
the label, 
the dimension, 
the number of grid cells,
the resolution of the best resolved region around the central star,
the radius $r_\mathrm{min}$ of the central sink cell, 
the initial mass $M_\mathrm{core}$ of the pre-stellar core, 
its corresponding free fall time 
$t_\mathrm{ff} = \pi ~ r_\mathrm{max}^{3/2} / \sqrt{8~G~M_\mathrm{core}}$, 
and the period $t_\mathrm{end}$ of evolution simulated are given. 
A '*' in the $t_\mathrm{end}$ column denotes that the whole
accretion phase of the star has been computed.
A '**' denotes that the computation has been stopped at the point in
time when no mass is left in the computational domain.
}
\end{deluxetable} 

\begin{deluxetable}{l c c c c c c c}
\tablecaption{
Overview of axially symmetric massive pre-stellar core collapse simulations
\label{tab:2Druns}}
\tablehead{
&
&
& 
\colhead{Resol.} &
\colhead{$r_\mathrm{min}$} &
\colhead{$M_\mathrm{core}$} &
\colhead{$t_\mathrm{ff}$} &
\colhead{$t_\mathrm{end}$} \\
\colhead{Label} &
\colhead{Dim.}  &
\colhead{Grid cells} & 
\colhead{(AU)} &
\colhead{(AU)} &
\colhead{$(\mbox{M}_\odot)$} &
\colhead{(kyr)} &
\colhead{(kyr)}
}
\startdata 
\multicolumn{2}{l}{2D Convergence runs} & \multicolumn{6}{l}{Sect.~\ref{sect:2DConvergence}} \\
\hline
2D-Convergence32x16  & 2D &  32 x 16    & 2.69 x 1.11 & 10   &  60 & 67.6 & 62\\
2D-Convergence64x4   & 2D &  64 x  4    & 1.27 x 4.18 & 10   &  60 & 67.6 & 93\\
2D-Convergence64x8   & 2D &  64 x  8    & 1.27 x 2.09 & 10   &  60 & 67.6 &
$691^{**}$\\ 
2D-Convergence64x16  & \multicolumn{7}{l}{see `2D-Mcore60Msol'} \\
2D-Convergence128x32 & 2D & 128 x 32    & 0.61 x 0.51 & 10   &  60 & 67.6 & $33^+$\\
\multicolumn{2}{l}{2D rmin runs} & \multicolumn{6}{l}{Sect.~\ref{sect:2Drmin}} \\
\hline
2D-rmin80AU  & 2D &  64 x 16 & 7.25 x 8.21 & 80  &  60 & 67.6 & $251^*$\\ 
2D-rmin10AU  & \multicolumn{7}{l}{see `2D-Mcore60Msol'} \\
2D-rmin5AU   & 2D &  64 x 16 & 0.69 x 0.52 &  5  &  60 & 67.6 & $631^+$\\ 
2D-rmin1AU   & 2D &  64 x 16 & 0.17 x 0.11 &  1  &  60 & 67.6 & $92^+$\\
\multicolumn{2}{l}{2D Mcore runs} & \multicolumn{6}{l}{Sect.~\ref{sect:2DMcore}} \\
\hline
2D-Mcore60Msol     & 2D &  64 x 16    & 1.27 x 1.04 & 10   &  60 & 67.6 &
$939^{**}$\\ 
2D-Mcore120Msol    & 2D &  64 x 16    & 1.27 x 1.04 & 10   & 120 & 47.8 &
$489^{**}$\\ 
2D-Mcore240Msol    & 2D &  64 x 16    & 1.27 x 1.04 & 10   & 240 & 33.8 &
$226^{**}$\\
% TODO: update no.
2D-Mcore480Msol    & 2D &  64 x 16    & 1.27 x 1.04 & 10   & 480 & 23.9 &
$41^+$\\
\enddata 
\tablecomments{
The table is structured in the same way as Table~\ref{tab:1Druns}.
Simulations, 
which are still running, 
are marked by an additional `$+$' in the $t_\mathrm{end}$ column.
}
\end{deluxetable} 

Aside from varying one specific parameter of the initial condition or the numerical configuration in each simulation
series, 
most of the initial conditions and the physics considered in the simulations stay the same. 

Our basic initial condition is very similar to the one used by 
\citet{Yorke:2002p1}.
We start from a cold ($T_0 = 20 \mbox{ K}$) pre-stellar core of gas and dust.
The initial dust to gas mass ratio is chosen to be 
$M_\mathrm{dust} / M_\mathrm{gas} = 1\%$.
The model describes a so-called quiescent collapse scenario without turbulent motion 
($\vec{u}_r = \vec{u}_\theta = 0$).
In non-spherically symmetric two-dimensional runs the core is initially in slow rigid rotation
$\left(|\vec{u}_\phi| / R = \Omega_0 = 5*10^{-13} \mbox{ Hz}\right)$.
The rotation speed of 
$\Omega_0$ 
results roughly in an equilibrium between gravity and centrifugal force at the outer core radius 
$r_\mathrm{max}$ 
in the case of the lowest mass core of 
$M_\mathrm{core} = 60 \mbox{ M}_\odot$. 
The outer radius of the cores is fixed to 
$r_\mathrm{max} = 0.1$ pc 
and the total mass
$M_\mathrm{core}$ 
varies in the simulations from 60 up to 480 $\mbox{M}_\odot$. 
The initial density slope drops with $r^{-2}$. 
The highest mass case of 480\Msol with a mean density of 
$\bar{\rho} \approx 8*10^{-18}\rhocgs~\left(\approx 2*10^6 cm^{-3}\right)$ denotes the upper mass limit 
of such a pre-stellar core we would expect from observations.
A brief overview of these physical initial conditions of the massive pre-stellar cores studied is given in 
Table~\ref{tab:InitialConditions}.

\begin{deluxetable}{c c l}
\tablewidth{0pt}
\tablecaption{
Overview of initial conditions
\label{tab:InitialConditions}}
\tablehead{
\colhead{Symbol} &
\colhead{Value}  &
\colhead{Quantity}
}
\startdata 
$T_0$ & 20 K & temperature of the pre-stellar core \\
$\left(M_\mathrm{dust} / M_\mathrm{gas}\right)_0$ & 1\% & dust to gas mass ratio
\\ $|\vec{u}_r|$ & 0 & radial velocity \\
$|\vec{u}_\theta|$ & 0 & polar velocity \\
$\Omega_0 = |\vec{u}_\phi| / R$ & $5*10^{-13}$ Hz & azimuthal angular velocity in 2D \\
$r_\mathrm{max}$ & 0.1 pc & outer radius of the pre-stellar core \\
$\rho(r)$ & $r^{-2}$ & density slope of the pre-stellar core \\
$M_\mathrm{core}$ & 60 to 480$\mbox{ M}_\odot$ & mass of the pre-stellar core \\
\enddata 
\end{deluxetable} 

The simulations are performed on a time independent grid in spherical coordinates (see Sect.~\ref{sect:discretization}).
The radially inner boundary of the computational domain is a semi-permeable wall towards the forming star, 
i.e.~the gas can enter the central sink cell, but it cannot leave. 
The outer radial boundary is a semi-permeable wall as well.
The mass can be pushed out of the computational domain 
(by radiative forces) 
but no mass is allowed to enter the computational domain.
The semi-permeable outer boundary implies the assumption that the collapsing core is mostly isolated from its
large-scale environment.
This limits the extent of the potential mass reservoir for the forming massive star to the initially fixed mass of the
pre-stellar core $M_\mathrm{core}$. 

The remaining numerical parameters are determined in several simulation series.
The resolution of the computational domain, 
which is necessary to follow the radiation and fluid physics as well as its interactions, 
is determined in several so-called convergence runs,
see Sects.~\ref{sect:1DConvergence} and \ref{sect:2DConvergence} for non-rotating and rotating cores respectively.
The highest resolution of the non-uniform grid is chosen around the forming massive star,
afterwards the resolution decreases logarithmically in the radial outward direction.
The default resolution goes down to 
$(\Delta r)_\mathrm{min} = 0.08$~AU and 
$\left(\Delta r \mbox{ x } r~\Delta{\theta}\right)_\mathrm{min} = 1.27 \mbox { AU x } 1.04 \mbox{ AU}$ 
for the one- and two-dimensional simulations respectively.
The accurate size of the sink cell is determined in parameter scans presented in Sects.~\ref{sect:1Drmin} and
\ref{sect:2Drmin} for non-rotating and rotating cores respectively.
While we started the first two-dimensional simulations with a radius of the inner sink cell of 
$r_\mathrm{min} = 80$ AU 
analog to the case of `F60' in \citet{Yorke:2002p1} 
we experienced that it is necessary to shrink the size of the sink cell 
down to a value smaller than the distance from the dust sublimation front to
the forming massive star, at least from the point in time on at which the radiative force becomes a serious counterpart to the gravity.
Otherwise a huge sink cell give rise to an artificially high radiative feedback 
and therefore limits the stellar mass reached in the simulations dramatically.
The default radius of the sink cell is chosen to be
$r_\mathrm{min} = 1$ AU and $r_\mathrm{min} = 10$ AU 
for the one- and two-dimensional simulations respectively.
In axially symmetric (two-dimensional) runs, 
physical shear viscosity is used to maintain the accretion flow through the growing circumstellar disk. 
Therefore, 
we adopted the well-known $\alpha$-para\-metri\-za\-tion model for shear viscosity of standard disk theory
\citep{Shakura:1973p3060}.
We performed several simulations with varying normalization values for the physical $\alpha$-viscosity,
which yield the formation of a stable accretion disk for a range of $\alpha$-values from 0.1 up to 1.0.
Apart from these runs the normalization of the viscosity was fixed to be $\alpha = 0.3$ here. 
A theoretical estimation on the $\alpha$-values of massive accretion disks was presented in
\citet{Vaidya:2009p12873}.

In previous test runs, 
we studied several non-radiative and radiative physics. 
We performed isothermal and adiabatic collapse simulations 
as well as gray and frequency dependent radiation transport
with and without radiation pressure feedback from the star or the diffuse thermal radiation field.
In this paper, 
we confine ourselves to present only the most realistic runs 
including frequency dependent radiation transport 
as well as full radiative feedback.

\section{Spherically symmetric accretion} 
\label{sect:1D}
As mentioned in the introduction, 
spherically symmetric accretion onto a massive star is potentially stopped by its growing radiation pressure. 
First, 
we present our results of one-dimensional simulations used to fix the
numerical parameters of the setup, namely, 
the resolution of the computational grid (see Sect.~\ref{sect:1DConvergence})
and the radius of the inner sink cell (Sect.~\ref{sect:1Drmin}). 
Afterwards, 
we analyze simulations with varying initial core masses $M_\mathrm{core}$ to determine the upper mass limit, 
if such a limit exists, 
for our specific model 
(the chosen dust and stellar evolution model, 
the configuration of the hydrodynamics solver as well as the treatment of radiation transport). 
To re-perform these one-dimensional simulations on our own instead of just referring to 
\citet{Larson:1971p1210}, 
\citet{Kahn:1974p1200}, 
\citet{Yorke:1977p1358}, 
and \citet{Wolfire:1987p539} 
allows us to directly compare
the results found for spherically symmetric accretion flows with subsequent simulation results in higher dimensions.

\subsection{Parameter scan of the size of the sink cell: The influence of the dust sublimation front}
\label{sect:1Drmin}
\subsubsection{Simulations}
\begin{figure}[htbp]
\begin{center}
\includegraphics[width=\FigureWidth]{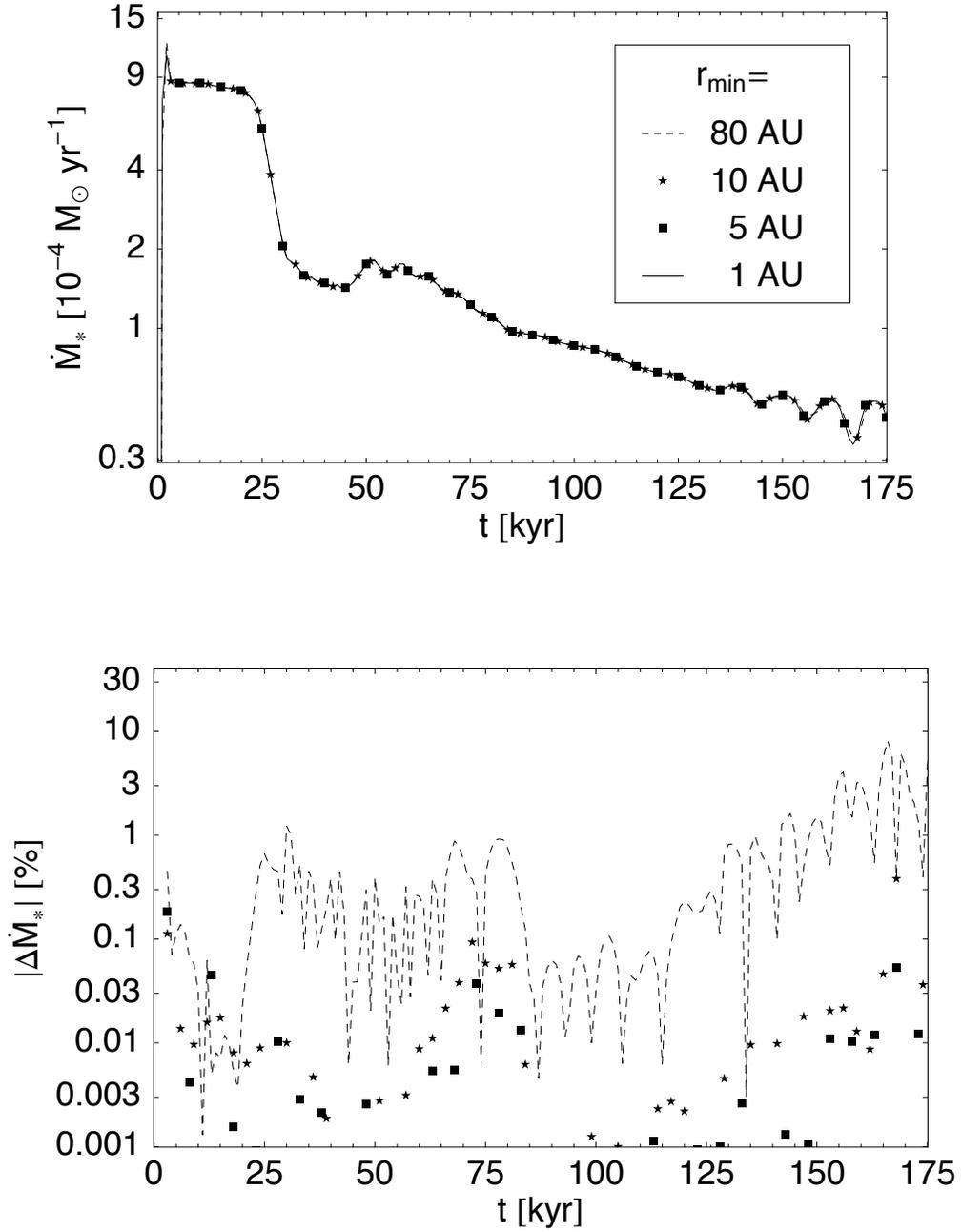}
\caption{
Accretion rate (upper panel) 
and deviations of the accretion rates from the simulation run with the
smallest sink cell radius of $r_\mathrm{min} = 1$~AU (lower panel) 
as a function of time for four different sizes of the
spherical sink cell. 
}
\label{1D_Rmin}
\end{center}
\end{figure}

In order to limit the run time of the simulations to an adequate amount, 
the formation and evolution of the central proto-star cannot be included in the computational domain.
In fact, 
the radially inner computational boundary defines the radius of a so-called sink cell.
The mass flux into this sink cell defines the accretion rate onto the proto-star, 
which is assumed to form in the center of the pre-stellar core.
Inside of this sink cell the stellar properties such as luminosity and radius are taken from
pre-calculated stellar evolutionary tracks.
We use therefore recent results for the evolution of accreting high-mass stars by
\citet{Hosokawa:2009p12591}.
In the following, 
we study the influence of the location $r_\mathrm{min}$ of this inner boundary on the resulting
accretion rate onto the evolving massive star.
We check this dependency in four simulations with a radius of the inner sink cell of $r_\mathrm{min} =$~1,~5,~10,~and~80~AU.
To decouple the results from the dependence on resolution 
(see Sect.~\ref{sect:1DConvergence}) 
the size of the grid cells was fixed to be
$\Delta r = 1$ AU up to a radius of 100 AU.
So the different simulations use 99, 95, 90, and 20 grid cells up to 100 AU respectively. 
Behind this inner region, 
the grid resolution decreases logarithmically throughout additional 128 grid cells from 100 AU up to 0.1 pc.
The initial conditions and numerical parameters of these runs are described in 
Sect.~\ref{sect:InitialConditions} 
and
the simulations are performed for an initial core mass of $M_\mathrm{core}=60\mbox{~M}_\odot$. 
We follow the long-term evolution of the runs for at least 200 kyr, 
representing 3.0 free fall times. 
The resulting accretion flow onto the forming star as well as the deviations
of the simulations from the run with the smallest sink cell
$r_\mathrm{min} = 1$~AU are displayed in Fig.~\ref{1D_Rmin}.

\subsubsection{Conclusions}
The first absorption of stellar irradiation takes place directly behind the dust sublimation radius $r_\mathrm{subl}$.
If the radius of the central sink cell $r_\mathrm{min}$ exceeds this dust sublimation radius, 
this interaction is artificially shifted to $r_\mathrm{min}$.
Due to the fact that the generalized Eddington limit is independent of the radius (the stellar gravity and the stellar
radiative flux both drop with $r^{-2}$), 
the shift of this first transfer of momentum from the stellar irradiation to the dust flow should
be independent of the radius of the sink cell.
Secondly, 
the absorption of stellar irradiation heats up the region behind the dust sublimation radius respectively.
The thermal radiative flux from this region outwards slows down the gravitationally in-falling accretion flow.
In general, 
this interaction depends on the radius, which defines the temperature of the heated region,
the velocity of the accretion flow and the opacity of the corresponding dust.

In the plot of the resulting accretion rates 
(Fig.~\ref{1D_Rmin}, upper panel) 
only slight deviations of the run with the largest sink cell radius 
$r_\mathrm{min} = 80$~AU 
are visible during the initial and final epoch. 
The other runs show identical results.
The lower panel of Fig.~\ref{1D_Rmin} shows in more detail the deviations of the simulations from the run
with $r_\mathrm{min} = 1$~AU. 
The simulation runs with $r_\mathrm{min} =$~10,~5,~and~1~AU stay identical.
In these simulations the dust sublimation radius,
which can be roughly estimated to 20~to~30~AU for a corresponding 20~to~30~$\mbox{M}_\odot$ star,
is included in the computational domain before the onset of radiation pressure occurs at roughly 25~kyr.
On the other hand, the largest sink cell of $r_\mathrm{min} = 80$~AU exceeds the dust sublimation radius
$r_\mathrm{subl}$.
The resulting accretion rate of the corresponding run oscillates around the results from the more precise simulations
with a maximum deviation of 10\% mostly at the end of the simulation, when the radiation pressure starts to revert the
accretion flow throughout the whole domain. 
Due to the fact that the deviations are oscillating and the fact that the strongest deviations occur at the end of
the simulation where the accretion rate is already an order of magnitude lower than at the beginning, 
the four simulations yield the same final mass of the proto-star.
Subsequent one-dimensional simulations presented make use of a radius of the central sink cell of 
$r_\mathrm{min} = 1$~AU.

\subsection{Parameter scan of the initial pre-stellar core mass: The upper mass limit of spherically symmetric
accretion} 
\label{sect:1DMcore}
\subsubsection{Simulations}
\begin{figure}[htbp]
\begin{center}
\includegraphics[width=\FigureWidth]{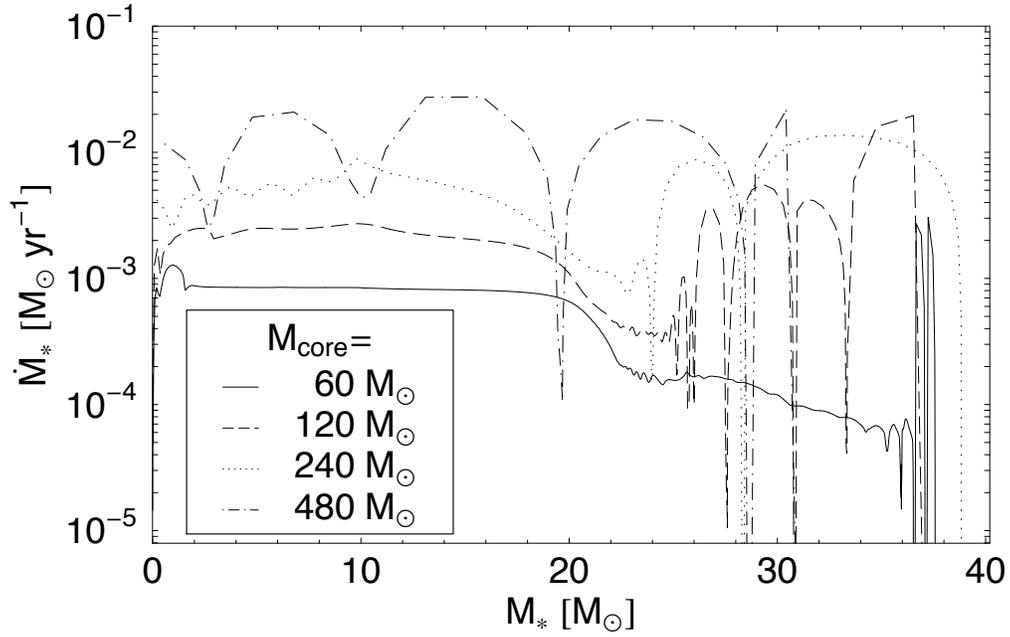}
\caption{
Accretion rate $\dot{M}_*$ as a function of the actual stellar mass $M_*$ for
four different initial pre-stellar core masses of $M_\mathrm{core} = 60 \mbox{ M}_\odot$ up to $480 \mbox{ M}_\odot$.
The spherically symmetric accretion models yield an upper mass limit of the final star of $M_*^\mathrm{1D} < 40 \mbox{
M}_\odot$. 
}
\label{1D_McoreScan}
\end{center}
\end{figure}

The simulations presented so far were performed to fix the remaining free
numerical parameters, 
namely, 
the grid resolution and the size of the central sink cell. 
We now study the collapse of massive pre-stellar cores for four different initial core masses
$M_\mathrm{core}$
ranging from 
$M_\mathrm{core} = 60 \mbox{ M}_\odot$ 
up to 
$480 \mbox{ M}_\odot$.
The initial conditions and numerical parameters for these runs are described in Sect.~\ref{sect:InitialConditions}.
The simulations are performed with an inner boundary of the computational domain of 
$r_\mathrm{min} = 1$~AU and 128 grid cells with logarithmically increasing resolution towards the center. 
The size of the innermost grid cell of the computational domain is 
$(\Delta r)_\mathrm{min} = 0.08$~AU.
We follow the evolution of the system until no mass is left in the computational domain. 
Part of this mass is accreted onto the central massive star and part is expelled over the outer boundary by
radiative forces.  
The resulting accretion histories are displayed in Fig.~\ref{1D_McoreScan} as a function of the actual stellar mass.

\subsubsection{Conclusions}
The mass and the luminosity of the forming massive star grow with time.
The radiation pressure of the direct stellar irradiation as well as from the thermal infrared dust emission increase
and
become stronger than gravity ultimately. 
Therefore, 
the accretion rate drops down and the massive star has grown to its final mass. 

\begin{figure}[htbp]
\begin{center}
\includegraphics[width=\FigureWidth]{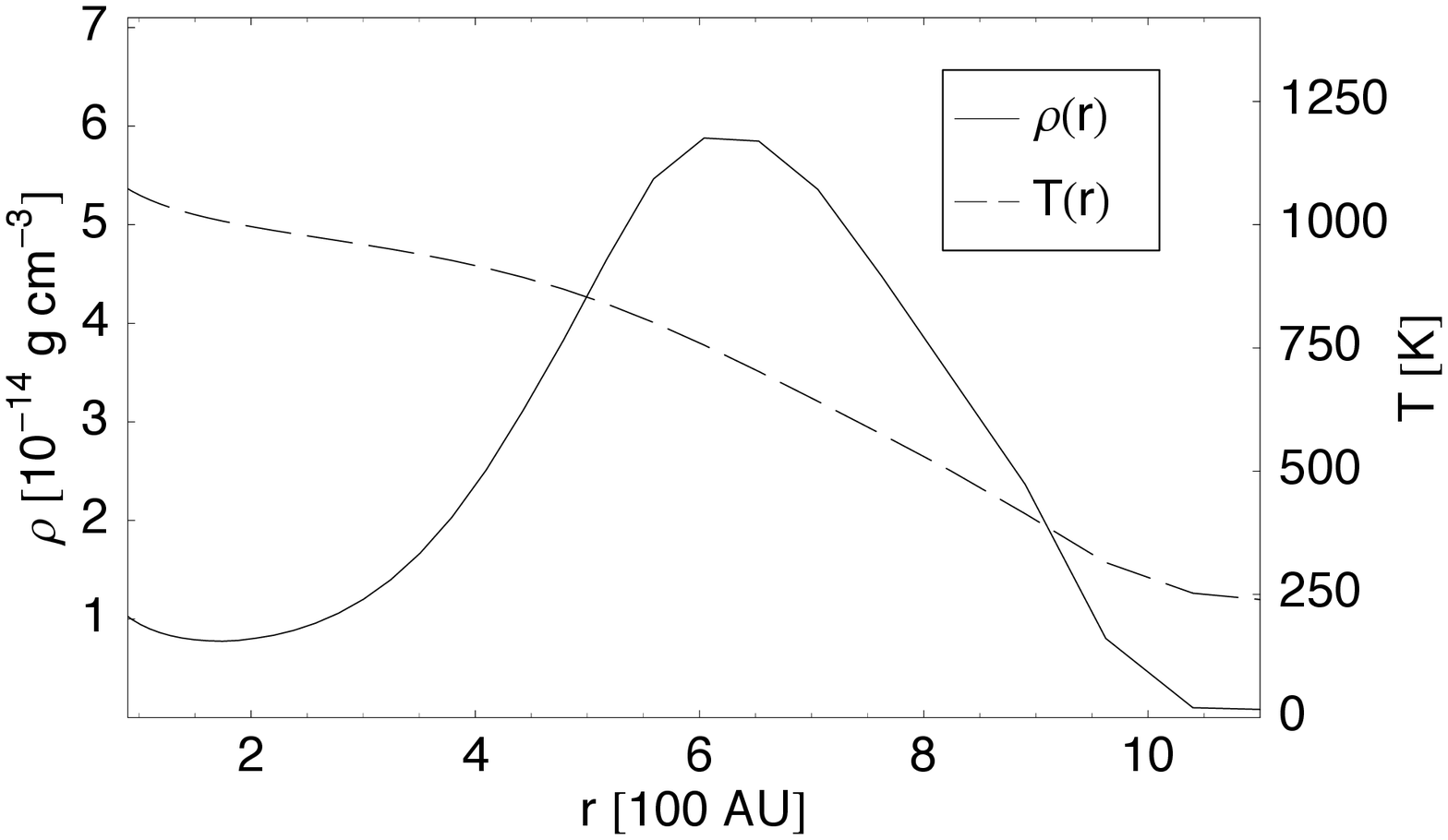}

\vspace{1cm}
\includegraphics[width=0.88\FigureWidth]{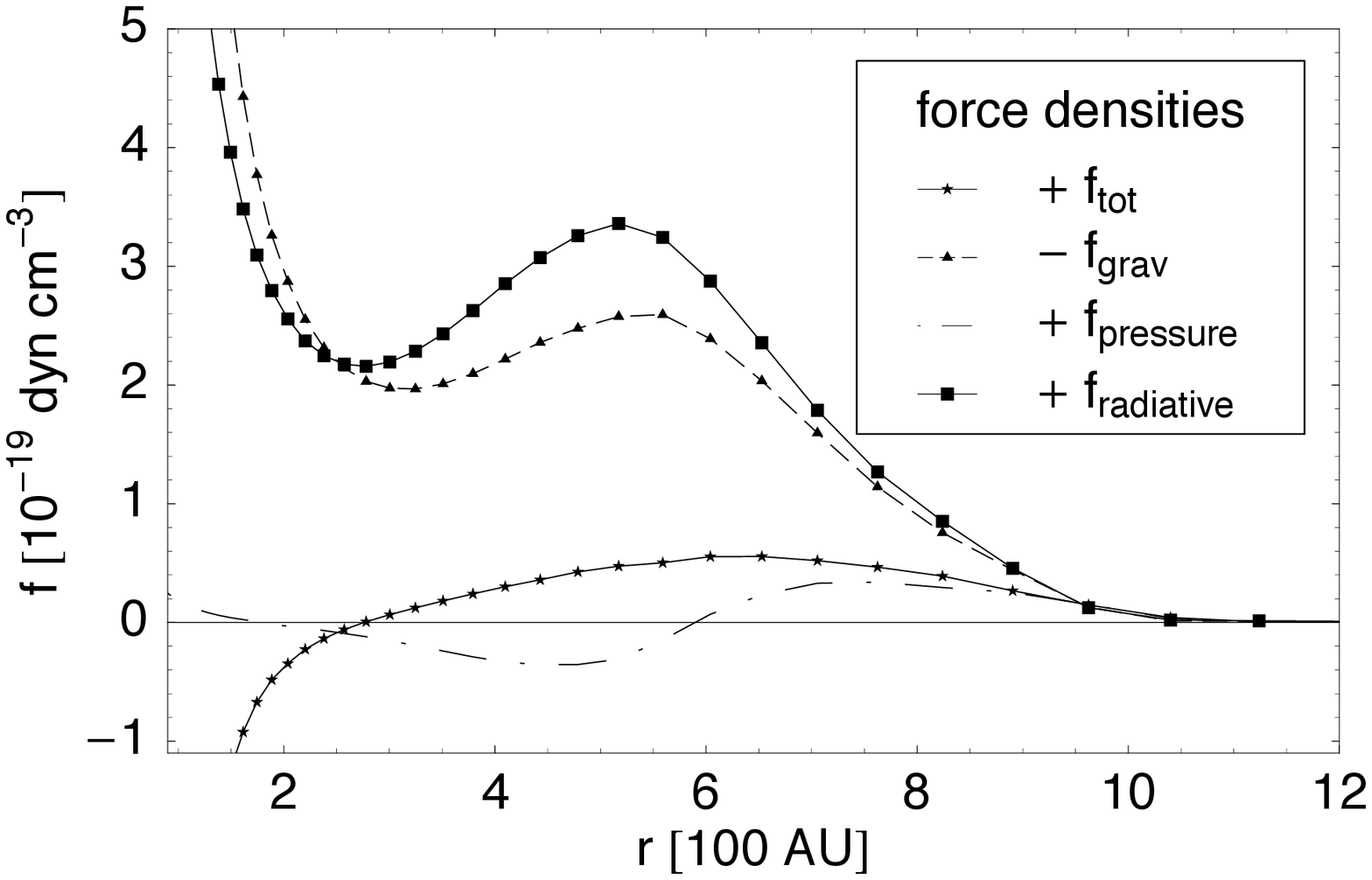}
\caption{
Snapshot of radial force densities (lower panel)
and the density and temperature profile (upper panel)
in the innermost core region taken from the collapse simulation of a 
$M_\mathrm{core} = 120 \mbox{ M}_\odot$ 
pre-stellar core at 20~kyr corresponding to a proto-stellar mass of about $M_* = 25 \mbox{ M}_\odot$.
Due to the superior radiative force the spherically symmetric accretion models yield an upper mass limit of the final
star of 
$M_*^\mathrm{1D} < 40 \mbox{ M}_\odot$.
}
\label{fig:1D_Accelerations}
\end{center}
\end{figure}

The individual force densities as as function of the radius through the spherically symmetric pre-stellar core are
displayed at a snapshot in time, where the radiative force starts to trigger the stopping of the in-fall motion in
Fig.~\ref{fig:1D_Accelerations}.
These forces are later on compared with the corresponding forces of the disk
accretion models.

The final star does not reach a mass higher than $40\Msol$ in any of the simulations.
This limit is in good agreement with previous research studies.
It lies in the allowed range of 25~-~60$\Msol$ determined by 
\citet{Larson:1971p1210}.
\citet{Kahn:1974p1200} predicted in his analytical study the formation of a $40\Msol$ star
and 
\citet{Yorke:1977p1358} formed a 36~$\mbox{M}_\odot$ star in their radiation hydrodynamics simulation of a $150\Msol$
collapsing core.

The oscillations of the accretion rate during the stopping of the in-fall motion are due to a negative feedback effect
of the accretion luminosity: 
By increasing the initial mass of the pre-stellar core from $60\Msol$ up to $480\Msol$, 
the amplitude of the accretion rate and therefore the accretion luminosity increases as well.
Due to the resulting stronger radiative force, 
the increase of accretion luminosity leads to a de-acceleration of the accretion flow, 
which results in a reduction of the corresponding accretion luminosity.
This negative feedback yields a highly episodic accretion history. 
The effect is stronger in cases,
where the ratio of the accretion luminosity to the stellar luminosity is high,
i.e. the effect is stronger for more massive cores.
Such an oscillating phase was also previously detected in the simulations by
\citet{Yorke:1977p1358}. 

The fact that the final mass of the star in the most massive case 
$M_\mathrm{core} = 480 \mbox{ M}_\odot$ is lower
($M_* \approx 31 \mbox{ M}_\odot$) 
than for the cores that initially had less mass, 
should be taken with care: 
In simulations with such high oscillations, 
the influence of the underlying stellar evolution model increases strongly.
To analyze the details of this time dependent interaction of the stellar evolution and the accretion flow, 
a self-consistent treatment of the proto-stellar's evolution and its environment should be considered.

\section{Disk accretion}
\label{sect:2D}
The most massive stars known cannot be formed by spherically symmetric accretion.
As shown in the last section, the radiative forces in a spherically symmetric envelope lead to a cut-off of the
accretion phase.
The high luminosity of a massive star heats the region in its vicinity to such
a high temperature that the resulting thermal radiation pressure overcomes the gravitational force.
The radiation pressure stops, and finally reverses the accretion flow.
Besides this theoretical issue, observations indicate the presence of angular momentum in all epochs of star
formation, starting with the rotation of pre-stellar cores
and finally resulting in rotating flattened circumstellar structures. 
Leaving perfectly spherical symmetry will thereby potentially help to overcome the radiation pressure problem.
First, the presence of higher densities in the forming disk region results in a thinner shell, where the first
absorption of stellar photons takes place.
This enables an accretion flow to break through this region of direct stellar
irradiation feedback more easily. 
Secondly, 
the feedback by radiation from dust grains, 
which actually stops the accretion in the spherically symmetric case, 
will be strongly reduced, 
because the majority of the radiative flux from the irradiated inner rim of the disk will escape in
the vertical direction through the optically thin disk atmosphere and therefore does not interact with the radially
inward-streaming accretion flow.
The different kinds of radiative feedback in spherical symmetry as well as in an axially symmetric disk geometry are
illustrated in the final discussion Sect.~\ref{sect:discussion}.

Analogously to the discussion of the spherical symmetric simulations, 
we present in the following the
results of axially and midplane symmetric simulations of the collapse of rotating massive pre-stellar cores.
Before being able to scan the parameter space of different initial core masses (Sect.~\ref{sect:2DMcore}), 
we determine the required resolution in convergence runs (Sect.~\ref{sect:2DConvergence}) 
and fix the radius $r_\mathrm{min}$ of the central sink cell 
in various simulations (Sect.~\ref{sect:2Drmin}). 

\subsection{Parameter scan of the size of the sink cell: The influence of the dust sublimation front}
\label{sect:2Drmin}
\subsubsection{Simulations}
\begin{figure}[htbp]
\begin{center}
\includegraphics[width=\FigureWidth]{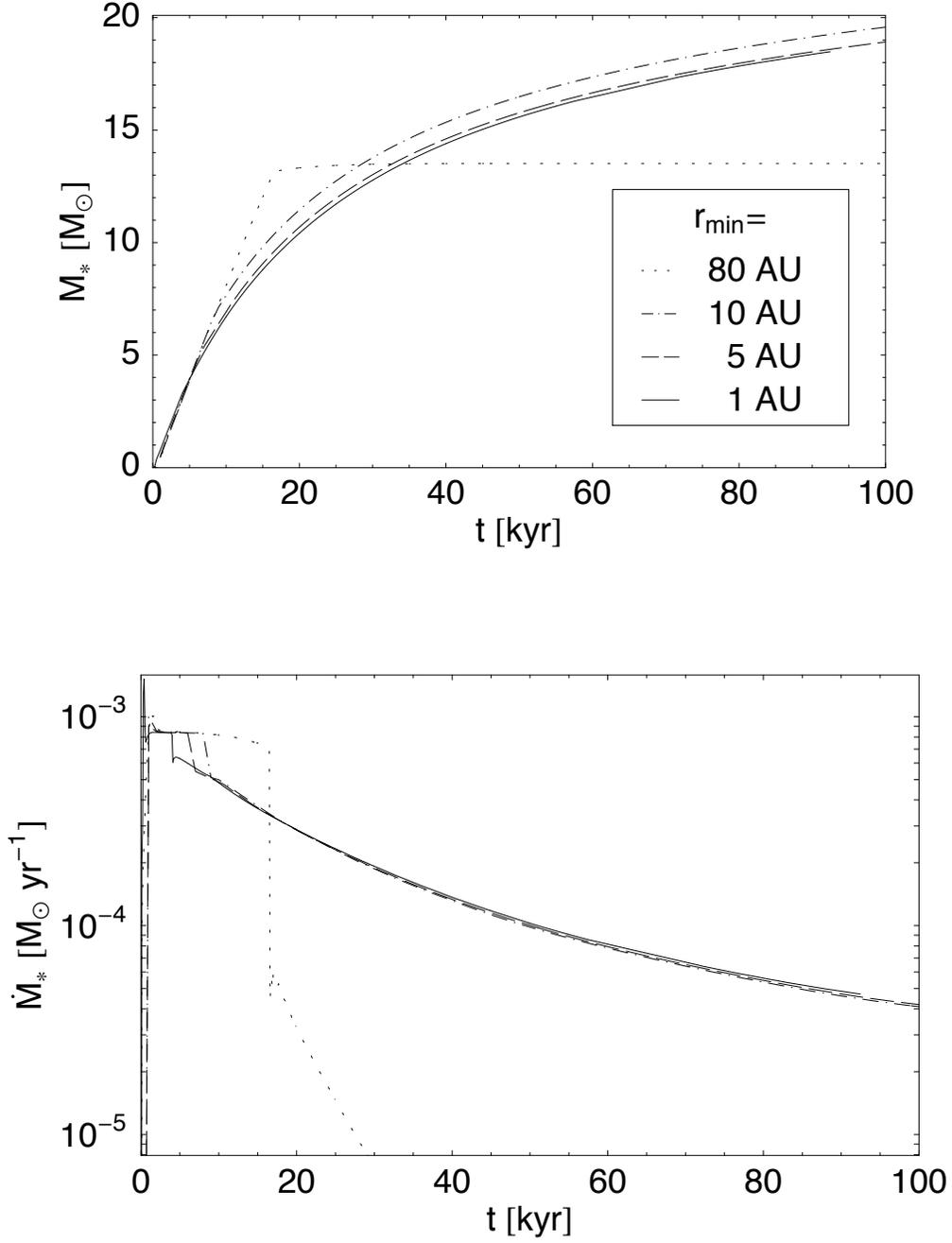}
\caption{
Stellar mass $M_*$ (upper panel) and accretion rate $\dot{M}_{*}$ (lower panel) as a function of time $t$ for different
radii $r_\mathrm{min}$ of the central sink cell in the collapse simulation of a rotating 60\Msol pre-stellar core. 
}
\label{fig:2D_Rmin_60Msol}
\end{center}
\end{figure}

In the following, 
we study the influence of the radius $r_\mathrm{min}$ of the inner sink cell, 
which equals the inner computational boundary, 
on the resulting accretion rate onto the evolving massive star.
We check this dependency in four simulations with a radius of the inner sink cell of 
$r_\mathrm{min} =$~1,~5,~10,~and~80~AU.
The initial conditions and numerical parameters of these runs are described in Sect.~\ref{sect:InitialConditions} 
and
the simulations are performed for an initial core mass of $M_\mathrm{core}=60\Msol$. 
%TODO: update numbers
We follow the long-term evolution of the runs for at least 92 kyr.
The resulting accretion rate onto the forming star as well as the mass growth
of the central star are displayed in Fig.~\ref{fig:2D_Rmin_60Msol}.

\subsubsection{Conclusions}
In the spherically symmetric models, we conclude that the numerical results do not depend on the radius $r_\mathrm{min}$ of the
central sink cell unless it is smaller than the dust sublimation radius $r_\mathrm{subl}$ from the point in time at which the
radiative force overcomes gravity.
These results cannot easily be transferred to the axially symmetric disk configuration.
Including centrifugal forces, which compensate the gravity in the disk region, the chosen location $r_\mathrm{min}$ of the
inner boundary of the computational domain influences the resulting accretion rate in two distinguishable effects:

Due to the fact that the circumstellar disk is growing in time from the inside outwards, a smaller sink cell leads to an
earlier onset of the disk formation phase during the simulation.
In other words, 
a fluid package with an initial position at $(r_i, \theta_i)$
and an initial rotation of $\Omega_i$ hold a centrifugal radius of
$r_\mathrm{cent} = \frac{\Omega_i^2 ~ r_i^4}{G ~ M(r_i)} \sin^2(\theta_i)$
with the included mass $M(r_i)$, see Eq.~\eqref{eq:includedmass}.
If this centrifugal radius is smaller than the sink cell radius $r_\mathrm{min}$, 
the fluid package is accreted onto the forming star during the
so-called free fall epoch at the beginning of the simulation.
This effect is associated with the gas physics (hydrodynamics) of the pre-stellar core, 
because the gas represents 99\% of the mass of the pre-stellar core.
A second important effect depending on the chosen sink cell radius is related to the dust and therefore to the radiation
physics.
The region in the vicinity of the forming massive star will be heated up to temperatures beyond the dust sublimation
temperature. 
Therefore a gap is formed between the central star and the dust disk.
Under the assumption that the absorption by gas in this gap is much smaller than
the absorption by dust grains behind the dust sublimation front the inner rim of the dust disk determines the region of the first stellar
radiative impact onto the accretion flow.
Also the most important radiative feedback due to thermal re-emission by dust grains sets in directly behind this
irradiated heated region. 
Contrary to the spherically symmetric case,
the high-density disk region bypasses most of the re-emitted radiation into the vertical direction, 
i.e.~the total radiation field is composed of the isotropic stellar irradiation and
a highly anisotropic (secondary) thermal radiation field.

Fig.~\ref{fig:2D_Rmin_60Msol} illustrates clearly both, 
the mass and the radiative effect, 
related to an artificial inner cut-off of the gas and the dust disk respectively:
As expected,
the duration of the so-called free fall phase shortens with the radius $r_\mathrm{min}$ of the sink cell.
This behaviour can fortunately be estimated analytically given the sink cell radius and the initial conditions of the
pre-stellar core to account for the overestimation of the final mass of the forming star, 
if necessary. 
Moreover, 
this effect of the artificial inner rim of the gas disk results 
on the one hand in an overestimation of the final mass of the central star by approximately 1\Msol or below 
(upper panel of Fig.~\ref{fig:2D_Rmin_60Msol}), 
but on the other hand influences the proceeding radiation hydrodynamic interactions in its environment only marginally 
(lower panel of Fig.~\ref{fig:2D_Rmin_60Msol}).
The corresponding accretion rates after the disk formation are not influenced at all.
This result is quite reasonable keeping in mind that the balance of radiative and gravitational forces
can be described in first order by the luminosity to mass ratio $L_\mathrm{tot} / M_{*}$ of the central massive star, 
which only changes marginally with another choice of the size of the central sink cell.

But the radiative impact due to the artificial cut-off of the inner dust disk regime for sink cell radii larger than
the dust sublimation front has dramatic effects.
In the case of $r_\mathrm{min} = 80$~AU the artificial shift of the region of
the dust radiation interactions terminates the short disk accretion phase abruptly,
leading to a completely wrong evolution of the central star, 
the disk as well as the large scale environment.
\vONE{
The reason for this dramatic change in the radiation physics is that the lower density region of the circumstellar disk at 80~AU is (in contrast to the real inner rim of the dust disk at roughly 20~AU) not opaque enough to generate a strong anisotropy of the thermal radiation field.
Therefore the strong isotropic part of the thermal radiation field is able to stop the accretion analogous to the spherically symmetric flow calculations.
}

\begin{figure}[htbp]
\begin{center}
\includegraphics[angle=270,width=\FigureWidth]{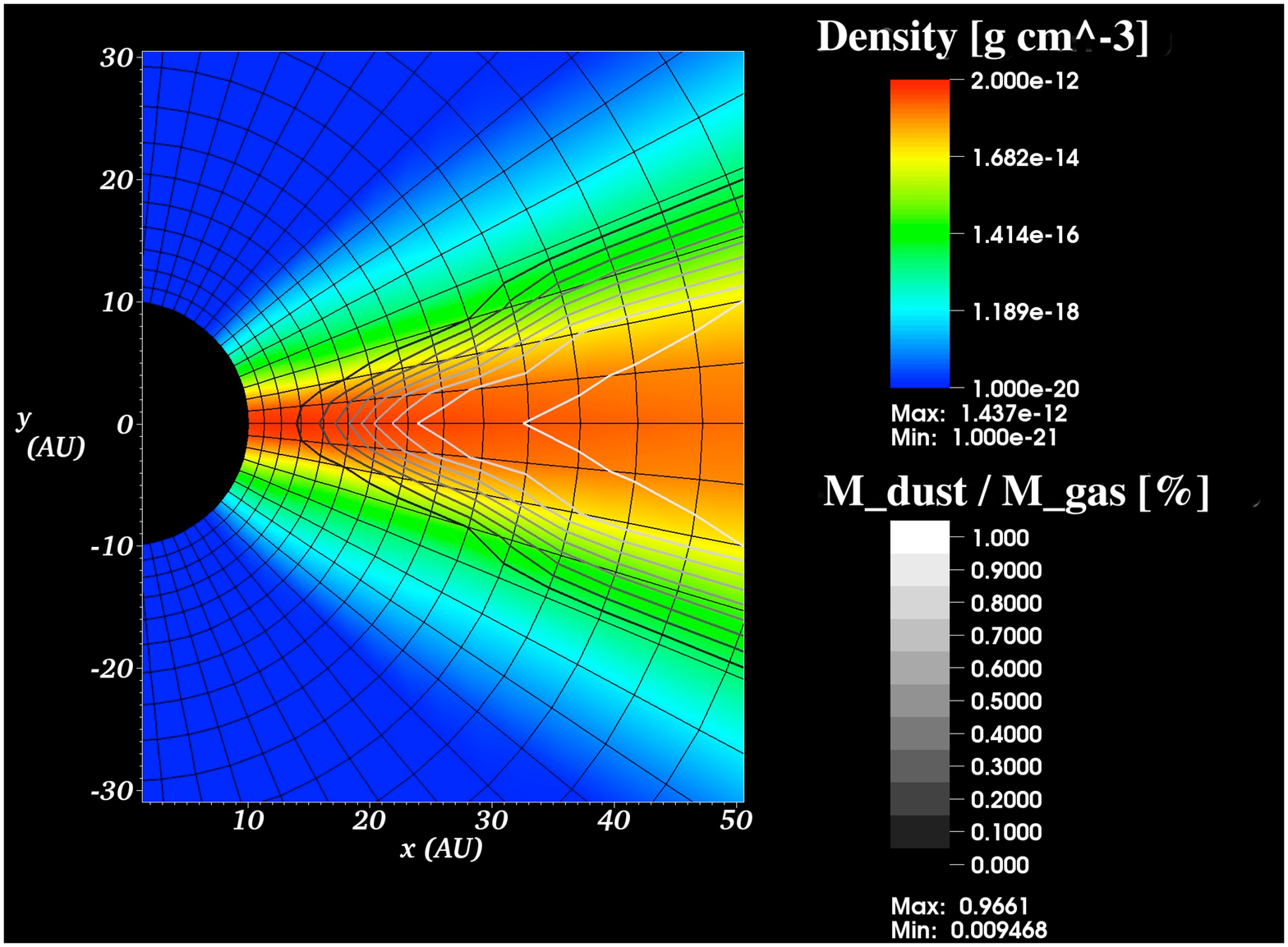}
\caption{
Resolving the dust sublimation front.
The image shows a snapshot at 100~kyr after start of the collapse of a 120\Msol pre-stellar core.
\newline
Color: Gas density from $10^{-20}$ up to $2*10^{-12} \rhocgs$ in logarithmic scale.
\newline
Contour-lines: Dust to gas mass ratio from 0 up to 1\% in linear scale.
\newline
The zoom-in illustrates the fact
that the dust sublimation front is resolved smoothly over several (here roughly 8) grid cells in the radial direction.
}
\label{fig:2D_DustCondensationFront}
\end{center}
\end{figure}

Due to the importance of this inner core region for the associated interaction of the radiation with the accretion flow
it seems to be unavoidable to include the whole dust disk down to its inner rim
in the computational domain (cp.~Fig.~\ref{fig:2D_DustCondensationFront}). 
This defines an upper limit of the radius
$r_\mathrm{min}$ of the central sink cell, which has to be smaller than the dust sublimation radius $r_\mathrm{subl}$ in the midplane from that point in time at which the radiative force has
grown to a competetive magnitude compared to the viscous force driving the accretion flow. 
Subsequent simulations meet this concern by using an adequate central sink cell radius of $r_\mathrm{min} = 10$~AU. 
Otherwise, 
for an inner sink cell radius $r_\mathrm{min}$ larger than the dust sublimation radius $r_\mathrm{subl}$, 
the region of radiative feedback is artificially shifted to higher radii including a strong decrease in 
density, opacity, and gravity.
The resulting strong heating of the disk region behind the radius $r_\mathrm{min} > r_\mathrm{subl}$, 
which `realistically' would be shielded from the stellar irradiation by the inner parts of the disk, 
leads to a diminishment of the shielding/shadowing property of the massive accretion disk.
In case the radiation field retained therefore its isotropic character in major parts,
the radiation pressure stops the emerging disk accretion phase,
similar to the spherically symmetric case.

This dependency of the radiation pressure on the radius of the sink cell explains also the abrupt end of the accretion
phase in the simulations by \citet{Yorke:2002p1}.
They presented simulations of collapsing pre-stellar cores of $M_\mathrm{core} = 30 \Msol$, $60 \mbox{
M}_\odot$ and $120 \mbox{ M}_\odot$.
The radius of their inner sink cell was chosen proportional to the initial mass of the core to be 40, 80, and 160~AU
respectively. 
As shown in this parameter scan 
(cp.~Fig.~\ref{fig:2D_Rmin_60Msol}) 
such huge sink cells lead to an artificial cut-off of the dust disk and result therefore in unphysically strong
radiative feedback. 
Therefore, 
we are definitely sure that this yields also the abrupt and early end of the accretion phase in the simulations by
\citet{Yorke:2002p1}. 

Contrary to our conclusion, 
\citet{Krumholz:2009p10975} stated that the much longer accretion phases in their own simulations
compared to \citet{Yorke:2002p1} are a result of non-axially symmetric modes in the outflow region. 
The physical argument in that case is that their simulations show a so-called ``3D radiative Rayleigh-Taylor
instability'' in the radiation pressure driven outflow, 
which results in further mass accretion onto the circumstellar disk from the bipolar direction
instead of a steady outflow feature.
In axially symmetric simulations the feeding of the disk would therefore not be possible.
Our axially symmetric collapse simulations,
presented in the following section,
show a stable radiation pressure driven outflow
and the forming circumstellar disk gains enough mass from the huge mass reservoir of the envelope 
to maintain its shielding property over several free fall times,
in fact over a longer period ever simulated in previous research studies.

\subsection{Parameter scan of the initial pre-stellar core mass: Breaking through the upper mass limit of
spherically symmetric accretion} 
\label{sect:2DMcore}
\subsubsection{Simulations}
\begin{figure}[htbp]
\begin{center}
\includegraphics[width=\FigureWidth]{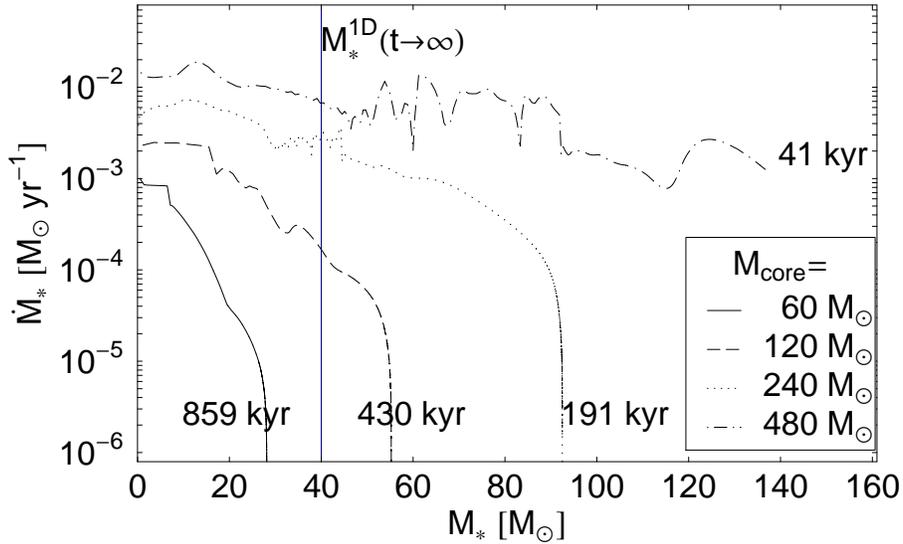}
\caption{
Accretion rate $\dot{M}_*$ as a function of the actual stellar mass $M_*$ for four different initial core masses 
$M_\mathrm{core} = 60 \mbox{ M}_\odot$, $120 \mbox{ M}_\odot$, 240$\Msol$, and $480 \mbox{ M}_\odot$.
Also the periods of accretion are mentioned for each run.
The vertical line marks the upper mass limit found in the spherically symmetric accretion models. 
The collapse models of slowly rotating pre-stellar cores clearly break
through this upper mass limit of the final star of 
$M_*^\mathrm{1D} < 40 \mbox{ M}_\odot$.
}
\label{fig:2D_McoreScan}
\end{center}
\end{figure}

The spherically symmetric (one-dimensional) collapse simulations of massive pre-stellar cores yield a maximum
stellar mass of less than 
$40 \mbox{ M}_\odot$ 
independent of the initial core mass 
$M_\mathrm{core} \geq 60\Msol$ 
due to radiative feedback. 
We attack this radiation pressure barrier 
in two-dimensional axially and midplane symmetric circumstellar disk geometry now.
The implications of this change of geometries are illustrated at full length in
the final discussion Sect.~\ref{sect:discussion}. 
We performed four simulations with the default initial conditions described in 
Sect.~\ref{sect:InitialConditions}
and the fixed numerical parameters presented in 
Sect.~\ref{sect:2DConvergence} and \ref{sect:2Drmin},
namely a maximum resolution of 1.27~AU~x~1.04~AU, 
and an inner sink cell radius of 10~AU.
The different initial core masses of 
$M_\mathrm{core} = 60 \mbox{ M}_\odot$,
$120 \mbox{ M}_\odot$, 240\Msol, and $480 \mbox{ M}_\odot$ are chosen analog to the scan of the
initial core mass parameter in the spherically symmetric case.
The resulting accretion histories as a function of the actual stellar mass are displayed in 
Fig.~\ref{fig:2D_McoreScan}.

\subsubsection{Conclusions}
As expected, 
the lowest mass case of $M_\mathrm{core} = 60 \mbox{ M}_\odot$ results finally in a less massive central star
than the corresponding run in spherical symmetry (without rotation) simply due
to the fact that the additional angular momentum results in centrifugal forces, 
which counteracts the accretion flow driven by gravity and viscosity. 
In face of this additional centrifugal forces,
for higher mass pre-stellar cores the slowed down accretion flux breaks easily through the upper mass limit of the
final star of 
$M_*^\mathrm{1D} < 40 \mbox{ M}_\odot$ 
found in the spherically symmetric accretion models! 

\begin{figure}[p]
\begin{center}
\includegraphics[width=0.88\FigureWidth]{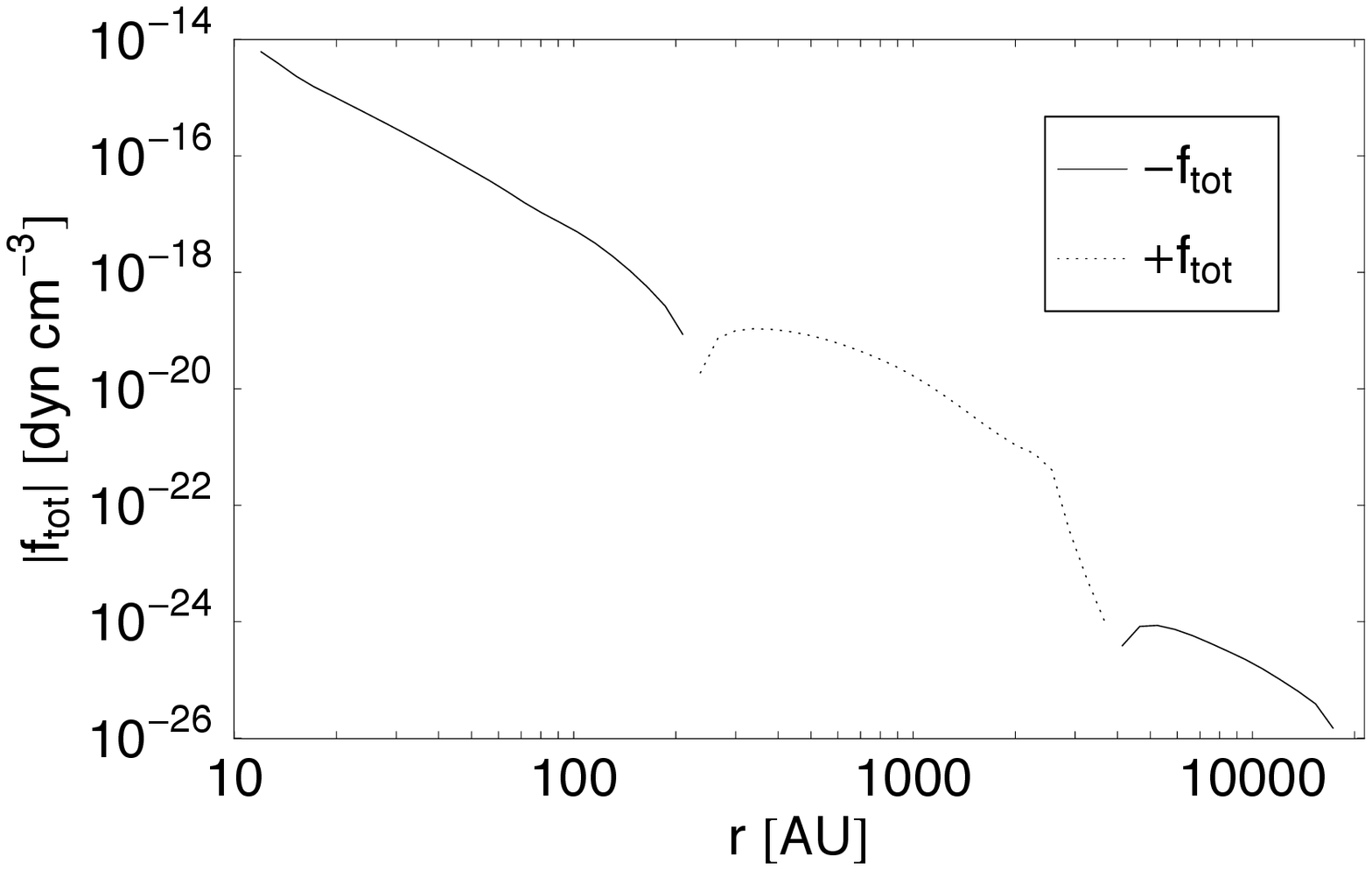}

\vspace{1cm}\hspace{2mm}
\includegraphics[width=\FigureWidth]{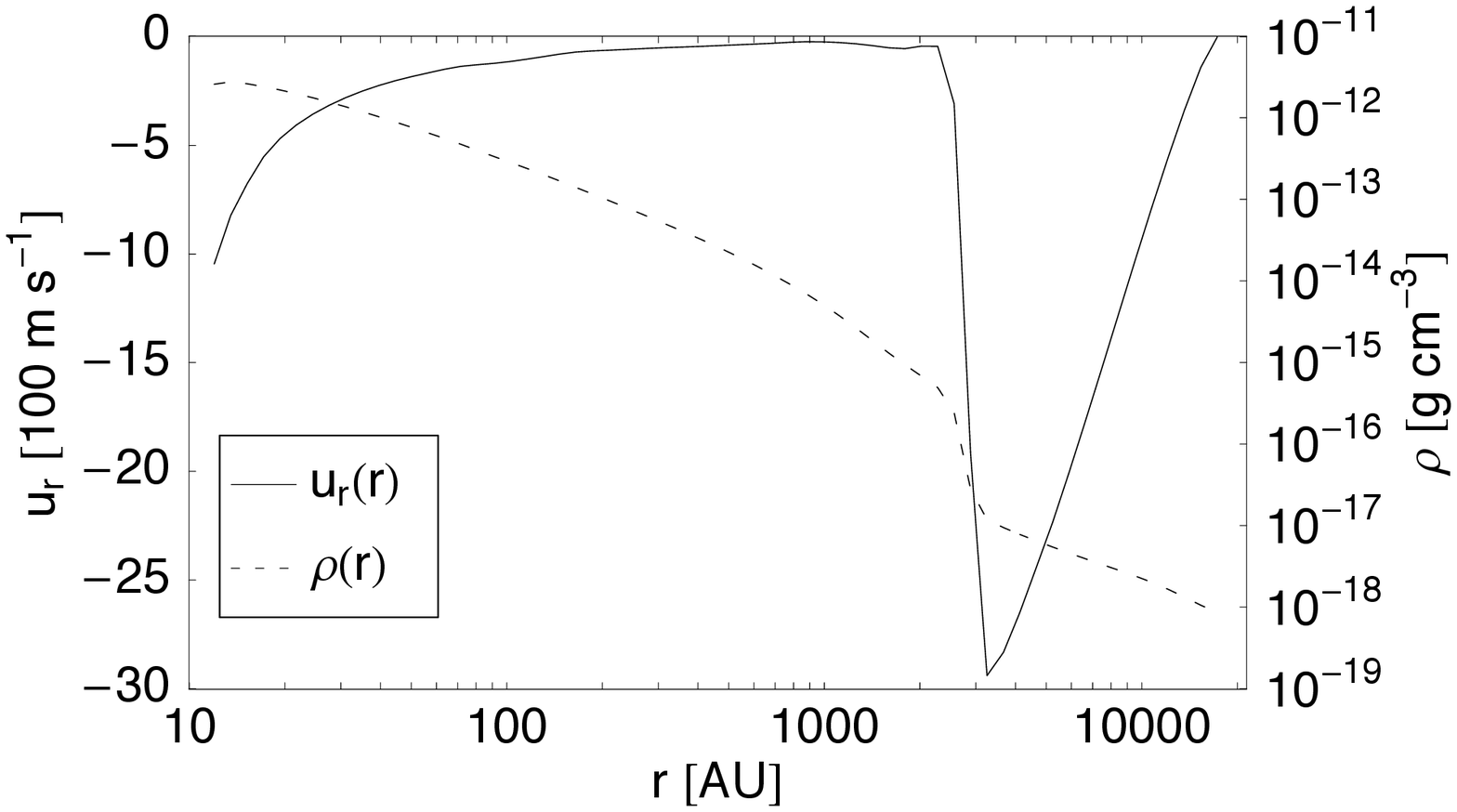}
\caption{
Total force density $|f_\mathrm{tot}(r)|$ (upper panel) 
as well as density $\rho(r)$ and radial velocity $u_r(r)$ (lower panel)
as a function of radius $r$ through the disk's midplane. 
The snapshot was taken at 60~kyr after start of the simulation, corresponding to a central stellar mass of roughly 
$40 \mbox{ M}_\odot$. 
The individual force densities along this line of sight through the total and the inner core region
are displayed in Figs.~\ref{2D_Accelerations_LargeScale} and \ref{2D_Accelerations_SmallScale}.
}
\label{2D_Accelerations}
\end{center}
\end{figure}

\begin{figure}[htbp]
\begin{center}
\includegraphics[width=0.75\FigureWidth]{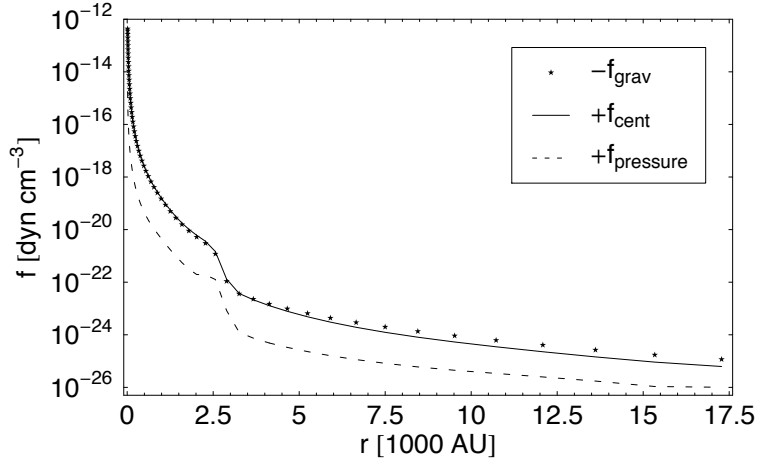}
\caption{
Gravity, centrifugal, and thermal pressure force as a function of radius through the disk's midplane.
The snapshot was taken at 60~kyr after start of the simulation, corresponding to a central stellar mass of roughly 
$40 \mbox{ M}_\odot$.
The radiative and viscous forces are orders of magnitude smaller than the illustrated ones, 
but become important in the inner disk region, 
where the stronger forces are in equilibrium. 
The radiative and viscous force densities along this
line of sight through the inner core region are displayed in Fig.~\ref{2D_Accelerations_SmallScale}.
}
\label{2D_Accelerations_LargeScale}
\end{center}
\end{figure}

\begin{figure}[htbp]
\begin{center}
\includegraphics[width=0.75\FigureWidth]{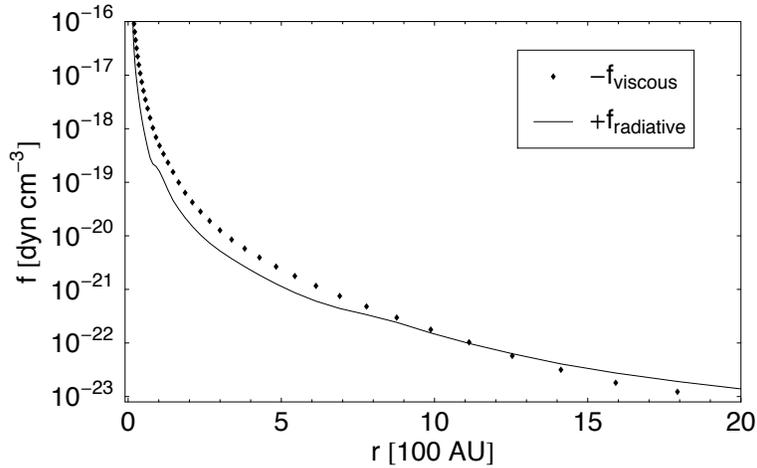}
\caption{
Viscous and radiative force density of the inner core region as a function of radius through the disk's midplane. 
The snapshot was taken at 60~kyr after start of the simulation, 
corresponding to a central stellar mass of roughly $40 \mbox{ M}_\odot$.
}
\label{2D_Accelerations_SmallScale}
\end{center}
\end{figure}

\begin{figure}[htbp]
\begin{center}
\includegraphics[width=0.88\FigureWidth]{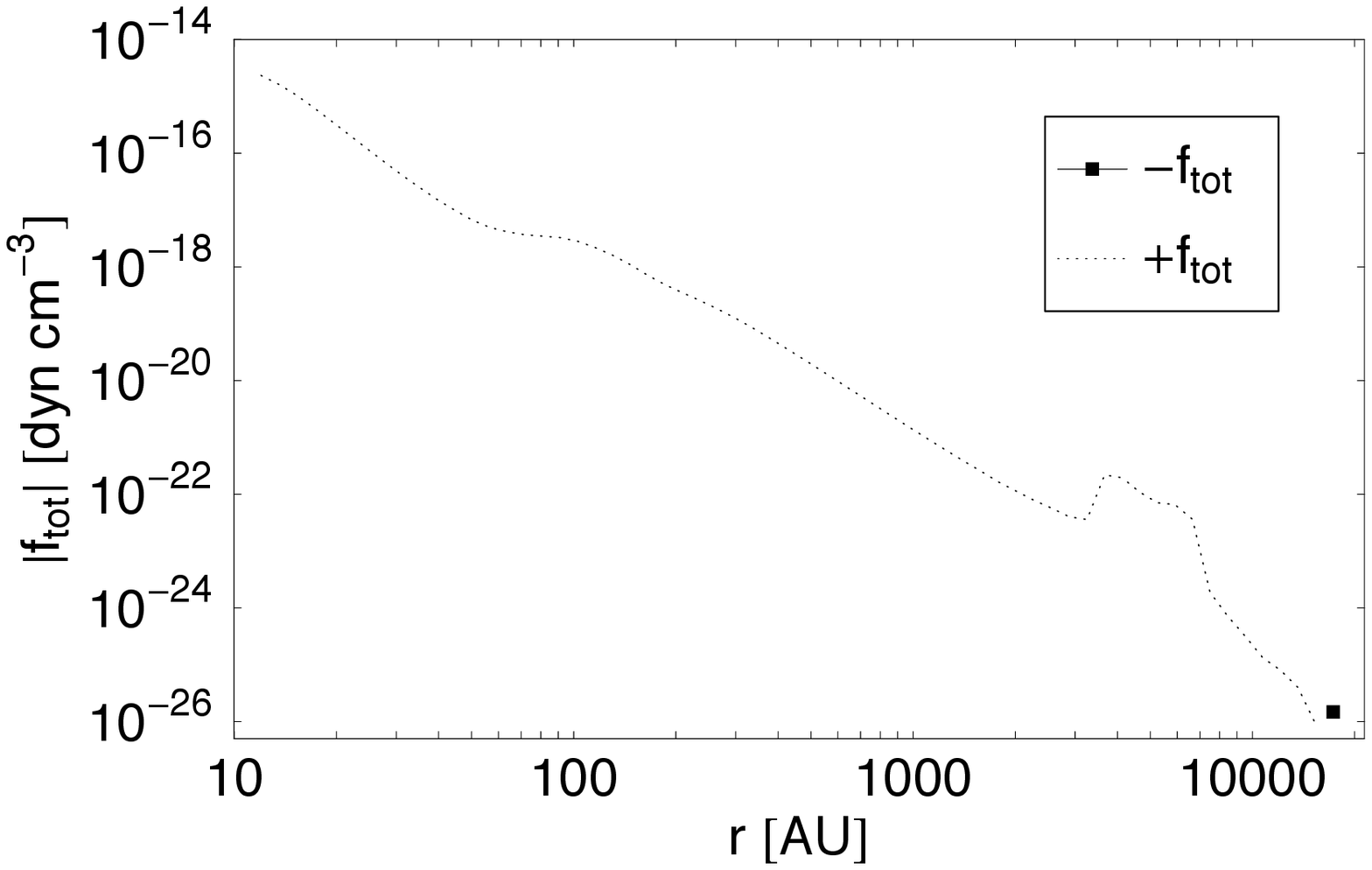}

\vspace{1cm}\hspace{2mm}
\includegraphics[width=\FigureWidth]{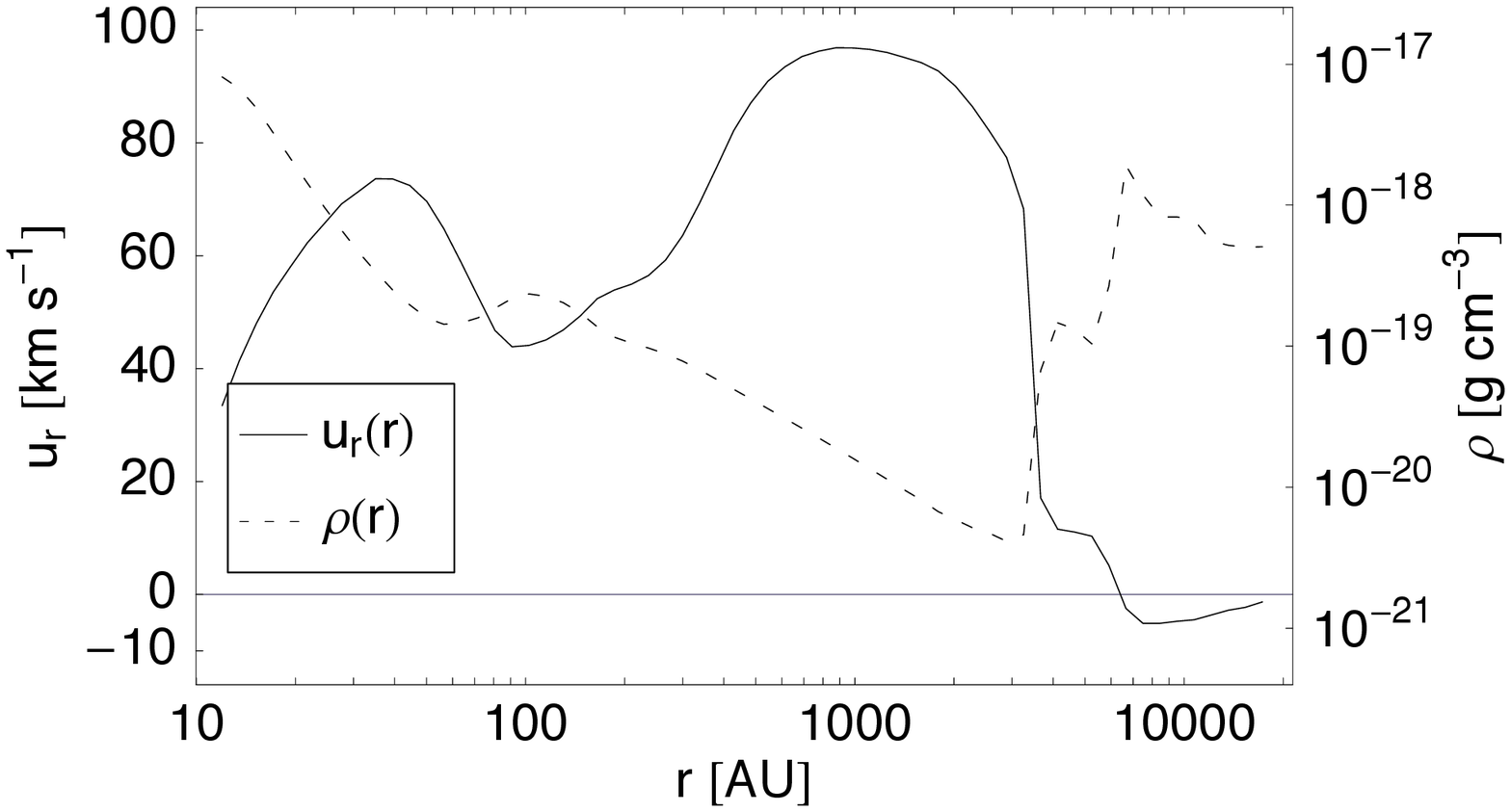}
\caption{
Total force density $|f_\mathrm{tot}(r)|$ (upper panel) 
as well as density $\rho(r)$ and radial velocity $u_r(r)$ (lower panel)
as a function of radius $r$ at $30\degr$ above the disk's midplane. 
The snapshot was taken at 60~kyr after start of the simulation, corresponding to a central stellar mass of roughly
$40 \mbox{ M}_\odot$. 
The individual force densities along this line of sight through the total and the inner core region are displayed in
Figs.~\protect\ref{2D_Accelerations_LargeScale_30deg} and \protect\ref{2D_Accelerations_SmallScale_30deg}.
}
\label{2D_Accelerations_30deg}
\end{center}
\end{figure}

\begin{figure}[htbp]
\begin{center}
\includegraphics[width=0.75\FigureWidth]{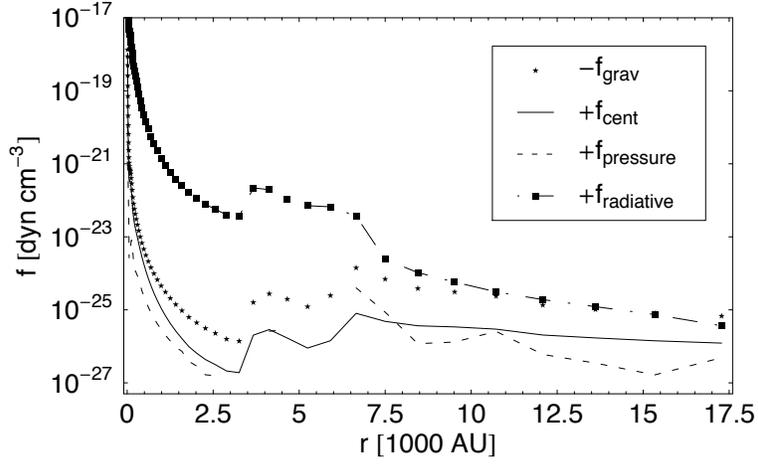}
\caption{
Gravity, centrifugal, thermal pressure, and radiative forces as a function of radius at $30\degr$ above the disk's
midplane. 
The snapshot was taken at 60~kyr after start of the simulation, corresponding to a central stellar mass of roughly
$40 \mbox{ M}_\odot$. 
The individual force densities along this line of sight through the inner core region are displayed in
Fig.~\ref{2D_Accelerations_SmallScale_30deg}.
}
\label{2D_Accelerations_LargeScale_30deg}
\end{center}
\end{figure}

\begin{figure}[htbp]
\begin{center}
\includegraphics[width=0.75\FigureWidth]{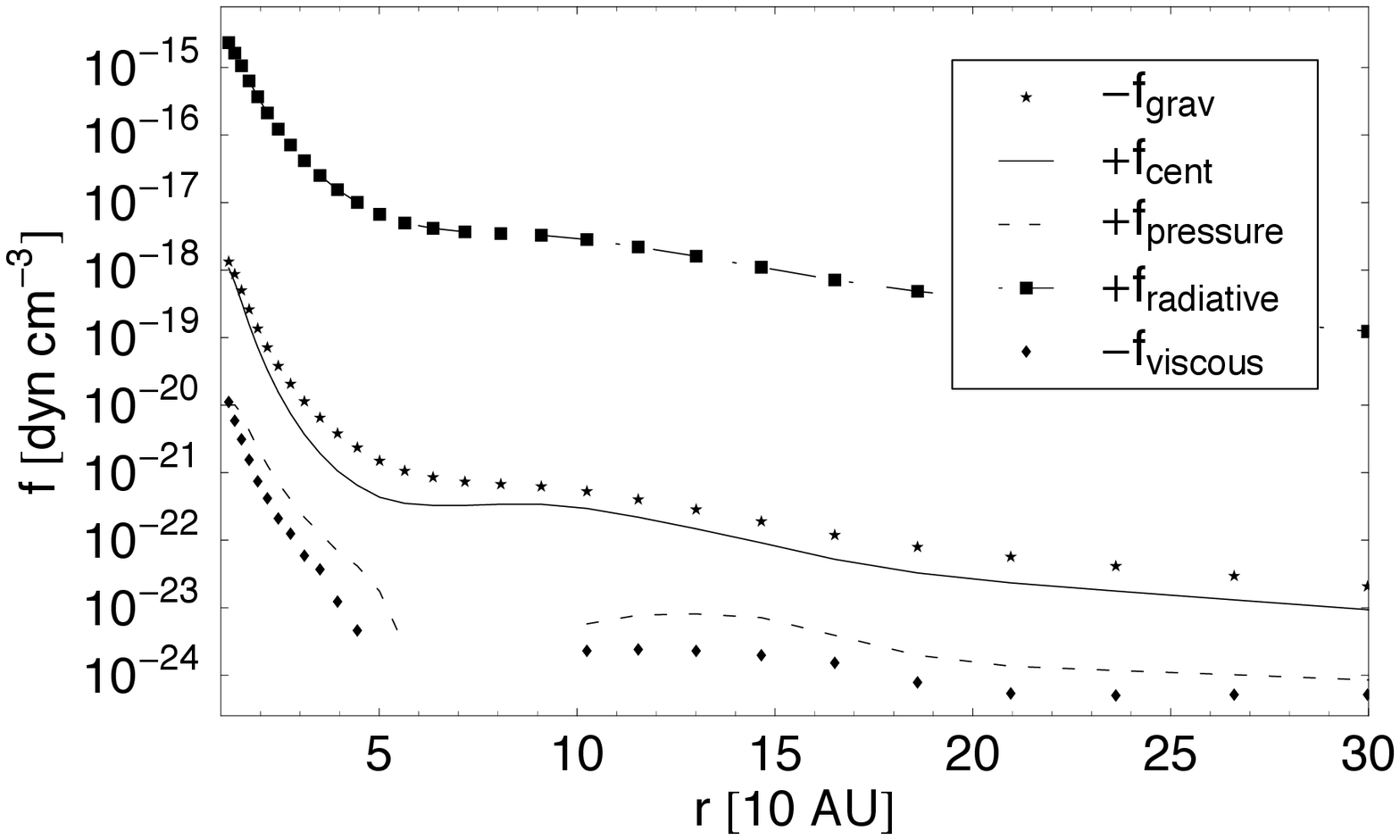}
\caption{
Gravity, centrifugal, thermal pressure, radiative, and viscous force density of the
inner core region as a function of radius at $30\degr$ above the disk's midplane. 
The snapshot was taken at 60~kyr after start of the simulation, corresponding to a central stellar mass of roughly 
$40 \mbox{ M}_\odot$.
}
\label{2D_Accelerations_SmallScale_30deg}
\end{center}
\end{figure}

The reason for that breakthrough can be displayed by a closer look at the driving
force densities in the evolved pre-stellar core, plotted in Figs.~\ref{2D_Accelerations} to
\ref{2D_Accelerations_SmallScale_30deg}.
All figures represent a snapshot of the 
$M_\mathrm{core} =$ 120$\Msol$ 
case at 60 kyr after start of the simulation.
At this point in time, 
the actual mass of the central massive star is roughly 40$\Msol$, 
representing the spherically symmetric
upper mass limit found in previous simulations (Sect.~\ref{sect:1DMcore}). 
In contrast to the spherically symmetric models, 
the geometry of the proto-stellar environment can now be divided
into a very dense circumstellar disk and the lower density envelope.
We visualized exemplary the actual density, velocity, 
and the acting forces in the radial direction for both regimes,
Figs.~\ref{2D_Accelerations} to \ref{2D_Accelerations_SmallScale} for the midplane of the accretion disk,
Figs.~\ref{2D_Accelerations_30deg} to \ref{2D_Accelerations_SmallScale_30deg} 
for a polar angle of $30\degr$ above the midplane.
In the midplane the gravity and centrifugal force are one to two orders of magnitude higher 
than the thermal pressure
and up to three orders of magnitude higher than the radiative and viscous force.
The upper panel of Fig.~\ref{2D_Accelerations} shows three individual regions of the midplane layer, 
in between the sign of the total force density changes. 
The gravity dominates the individual forces for the outer core regions 
(above 3000~AU) 
leading to a steady accretion flow onto the inner core region 
(Figs.~\ref{2D_Accelerations} and \ref{2D_Accelerations_LargeScale}).
In the very inner part of the core around the massive star 
(below 200~AU) 
the gravity is balanced by the centrifugal
force and in small part by the thermal pressure (Fig.~\ref{2D_Accelerations_LargeScale}).
In this region, 
which we will refer to as the disk region hereafter, 
the shear viscosity yields a quasi-stationary accretion flow through the disk, 
which clearly exceeds the radiative force 
(Fig.~\ref{2D_Accelerations_SmallScale}).
In between this disk region 
($<200$~AU) 
and the global in-fall region 
($>3000$~AU) 
the mass flux describes transient oscillations, 
because gravity, centrifugal forces and thermal pressure are not in equilibrium yet, 
as it is the case for the mass finally arriving the disk region.
Although the total force density, 
displayed in the upper panel of Fig.~\ref{2D_Accelerations} 
is directed in the outward direction between
$200 \mbox{ and } 3000 \mbox{ AU}$,
the mass in this region is still in an inward motion 
(cp.~Fig.~\ref{2D_Accelerations}, lower panel),
i.e.~the mass flow through the pre-stellar core is feeding the circumstellar accretion disk.
The viscous force in the accretion disk is able to drive a steady accretion flow 
towards the evolving massive star of 40\Msol, 
because the radiative force is one to two orders of magnitude lower in this dense disk region than in
the low density envelope (cp.~Figs.~\ref{2D_Accelerations_SmallScale} and \ref{2D_Accelerations_SmallScale_30deg}).
Observations of such a large scale flattened structure with a potentially embedded small scale accretion disk are
i.e.~described in
\citet{Fallscheer:2009p12914} and \citet{Beuther:2009p12913}.

At an polar angle of $30\degr$ above the midplane this strong radiative force already accelerates the remnant mass in
the radially outward direction through mostly the entire pre-stellar core
(Fig.~\ref{2D_Accelerations_LargeScale_30deg}). 
Only at the outer rim of the core the previous in-fall motion is still visible.
This distribution of the individual force densities confirms in high detail the
expected result of the anisotropy of the thermal radiation field.
Most of the radiative flux from the irradiated inner rim of the disk is bypassed in the vertical direction
through the optically thin atmosphere of the circumstellar disk.
Meanwhile, 
the accretion flow is reduced compared to the one-dimensional gravitational in-fall to a steady
stream driven by the viscous properties of the accretion disk.
In the envelope region of the pre-stellar core the radiative force reverts the in-fall motion and depletes the stellar
surrounding similar to the spherically symmetric accretion models 
(cp.~the corresponding individual force distributions in 
Figs.~\ref{2D_Accelerations_30deg} and \ref{fig:1D_Accelerations}).

In these axially and midplane symmetric disk accretion models no upper mass limit of the final star is detected so far,
but the star formation efficiency declines for higher mass cores. 
The proceeding depletion of the envelope by radiative forces finally leads to a decrease of the density in the
midplane and the disk looses its shielding property.
Without this shielding the radiation pressure starts to accelerate the remnant material in the outward direction.

\section{Discussion}
\label{sect:discussion}
\subsection{Radiation pressure feedback in a nutshell}
\begin{figure}[htbp]
\begin{center}
\includegraphics[angle=270,width=\FigureWidth]{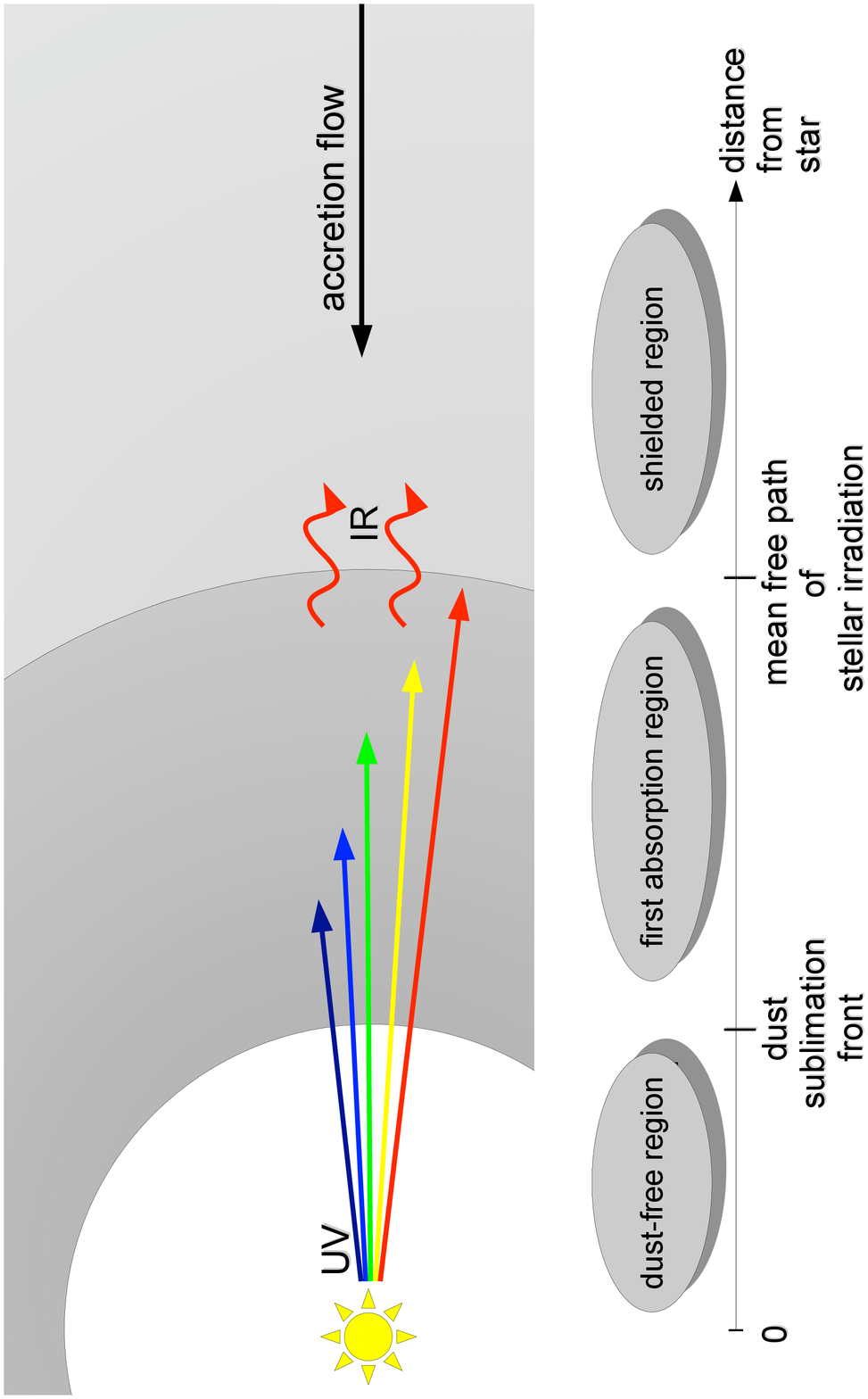}
\caption{
Schematic view of the radiative forces onto the accretion flow in spherical symmetry. 
The radiative feedback is divided into direct stellar irradiation and secondary re-emitted photons by dust grains. 
}
\label{1D_RadiativeForce}
\end{center}
\end{figure}

Since the isotropic and anisotropic feature of the spherical and disk accretion scenario at the dust
sublimation front is of such a great importance in our simulation results,
we illustrate these key attributes in more detail:
In this research study, 
we investigated the influence of the stellar environment 
onto the radiation pressure problem in the formation of massive stars.
We studied the accretion flow onto a high-mass star in a monolithic pre-stellar core collapse picture, 
as recommended by
\citet{Whitney:2005p5067} and \citet{McKee:2007p848}. 
Under this assumption, 
the theoretical description of the accretion process onto a massive star has to deal with the interaction 
between the exerted radiation by the forming star with the accretion flow of gas and dust 
\citep{Shu:1987p1616}. 
In a perfectly spherically symmetric collapse, 
this interaction potentially stops the accretion onto the star entirely. 
In the static limit, 
radiation pressure overcomes gravity at the so-called generalized Eddington barrier 
\begin{equation}
\frac{L_*}{M_*} = \frac{4\pi~G~c}{\kappa_*},
\end{equation}
where
$L_*$, $M_*$ and $\kappa_*$ denote the stellar luminosity, the stellar mass, and the dust opacity respectively,
$G$ is the gravitational constant and $c$ is the speed of light. 
But the collapse of a pre-stellar core is far from being a static problem. 
The momentum transfer from the absorbed photons first has to slow down the in-falling envelope.
For simplification purposes, 
we can divide the radiation pressure feedback into two components,
as illustrated in Fig.~\ref{1D_RadiativeForce}. 
The first exchange of momentum takes place when the irradiation from the massive star is absorbed by the dust
grains of the surrounding, 
i.e.~behind the dust sublimation radius. 
The strongest force will thereby be produced by photons with shorter wavelengths, 
because they have a higher absorption probability and are more energetic.
We will call this first interaction ``UV feedback'' abbreviated, 
although the frequency dependence of the broad stellar black body spectrum is clearly not negligible.
Afterwards, 
these heated regions emit most of the photons at the dust temperature,
yielding a much longer wavelength and a much longer mean free path than the direct stellar light. 
The interaction of this radiation with the enclosed gas and dust is therefore referred to as ``IR feedback''.
Our spherically symmetric collapse simulations confirm the outcome of previous
studies that it is essentially the IR feedback that stops the accretion flow onto the forming star in spherical symmetry. 
Although each IR photon transfers less momentum to the dust than the highly energetic stellar UV photons, 
the thermal dust emission acts onto the accretion flow on a much larger volume 
than the spatially confined absorption region of the stellar irradiation.
Additionally the optical depth of the envelope decreases towards \vONE{longer} wavelength,
so the IR feedback conteracts the accretion flow in the outer core regions yielding less gravity.
Different approaches to overcome this barrier for spherically symmetric accretion flows onto massive stars were
considered in the past. 
The generalized Eddington barrier depends only on the stellar evolution 
$\left(L_*/M_*\right)$ 
and on the dust properties 
$\left(\kappa_*\right)$.
\citet{Wolfire:1987p539} studied the necessary change of dust properties to enable further accretion, 
but the restrictions they derived seem to be unrealistic.

\begin{figure}[htbp]
\begin{center}
\includegraphics[angle=270,width=\FigureWidth]{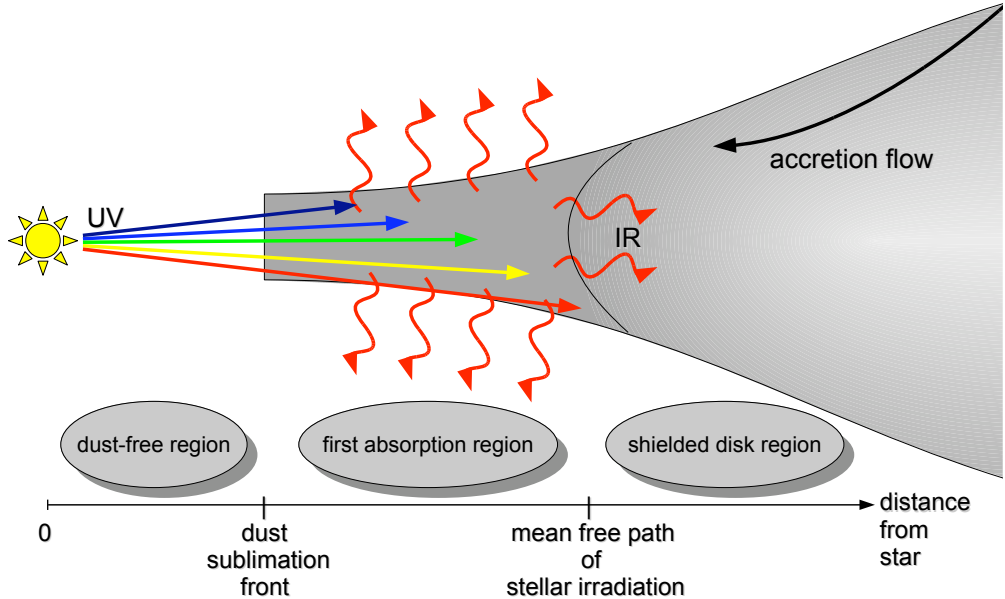}
\caption{
Schematic view of the ``UV"- and ``IR"-component of the radiation pressure 
acting in an axially symmetric circumstellar disk geometry.
}
\label{2D_RadiativeForce}
\end{center}
\end{figure}

Without a doubt,
star formation is rarely a perfectly spherically symmetric problem. 
Initial angular momentum of the collapsing pre-stellar core leads to the formation of a circumstellar disk 
as well as polar cavities. 
Compared to the case of spherically symmetric accretion,
the disk geometry changes the radiation pressure feedback dramatically, 
see Fig.~\ref{2D_RadiativeForce}.
Going from a spherically symmetric in-fall to an axially symmetric disk geometry can help to overcome both 
- the UV as well as the IR - 
radiation pressure feedback: 
Developing radiation hydrodynamical
\citep{Klahr:2003p2794},
magneto-rotational
\citep{Balbus:1991p3799, Hawley:1991p11900, Balbus:2003p11906}
and self-gravitating instabilities 
\citep{Yang:1991p11485, Laughlin:1994p11930, Bodenheimer:1995p11927}
in the accretion disk will transfer angular momentum to outer radii allowing a steady mass accretion radially inward. 
The additional ram pressure in the radiatively shielded parts of the disk will possibly push the mass over the thin
shell of the UV feedback 
\citep{Nakano:1989p1267}. 
Secondly, and most important, 
the irradiated and therefore heated regions of the disk will mainly cool in the vertical direction through the
optically thin disk's atmosphere, 
strongly restraining the IR radiation pressure in the radial direction. 
If the latter process occurs at the innermost part of the disk so that the radiation can escape directly through the
bipolar cavity, 
this effect is also known as the so-called `flashlight-effect' 
\citep{Yorke:2002p1, Krumholz:2005p867}.

The interaction of the radiation with the accretion flow is very sensitive to the numerical treatment of radiation transport. 
The FLD approximation, which is a standard technique in modern radiation hydrodynamics codes for astrophysical fluid flows, fails to compute the correct flux between the first transition region from the dust depleted zone around the massive star and the optically thick disk leading to an incorrect temperature distribution in the irradiated regions \citep[see~e.g.][]{Yorke:1977p1358, Boley:2007p2959}.
Also simplifying the stellar black body spectrum by using frequency averaged Planck mean opacities leads to a thinner shell of the direct stellar irradiation feedback and a stronger heating of the corresponding dust, which afterwards yields a higher IR feedback. 
Hence, accounting for the frequency dependence of the stellar spectrum seems to be a crucial point. 
%ONEHaving in mind the strong influence of the correct treatment of the stellar irradiation, we developed a radiation transport module for three-dimensional hydrodynamics simulations which includes a fast gray FLD solver for the appropriate dust cooling as well as a frequency dependent first order ray-tracing routine for the careful treatment of the stellar irradiation \citep{Kuiper:2010p12874}. 
%ONEBoosted by a modern solver library for large but sparse linear systems of equations \citep{petsc-web-page}, the radiation transport (as well as the self-gravity problem) can be solved quickly and accurately on distributed memory machines.

The most violent interaction of the stellar irradiation with the accretion flow takes place at and directly behind the first absorption peak. 
The location of the first absorption layer is represented by the dust sublimation front,  where the local dust temperature falls below the evaporation temperature of the dust grains. 
A systematic study of the radiation pressure feedback on the formation of massive stars therefore implies the need to resolve the ongoing radiation and accretion physics down to the dust sublimation front. 
A formation of massive stars by breaking through the ionization boundary into regions of sublimated dust grains was studied for spherically symmetric accretion flows \citep{Keto:2003p1242} as well as for two-dimensional effects in the so-called small radius limit \citep{Jijina:1996p1212}.
Aside from the important contribution of the proceeding physics at the dust sublimation front, no previous non-spherically symmetric numerical research has been done so far, presumably due to resolution issues.

\subsection{Comparison to previous numerical research in the field}
In our simulations the star in the center of the accretion disk grows far beyond the upper mass limit found in the case of spherical accretion. 
Indeed, the final massive stars are the most massive stars ever formed in a multi-dimensional radiation hydrodynamics simulation so far. 
The quantitative final results as well as a comparison to previous multi-dimensional radiation hydrodynamics studies is presented in Table~\ref{tab:summary}.
\begin{deluxetable}{l c c c c}
\tablecaption{
Overview of multi-dimensional radiation hydrodynamics simulations of massive star formation
\label{tab:summary}}
\tablehead{
\colhead{} &
\colhead{$M_\mathrm{core}$} &
\colhead{$t_\mathrm{end}$} & 
\colhead{$M_*$} &
\colhead{SFE} \\
\colhead{Authors} &
\colhead{(\Msol)} &
\colhead{(kyr)} & 
\colhead{(\Msol)} &
\colhead{(\%)}
}
\startdata 
       &  30 & 25 & 31.6 & \nodata\\
\protect\citet{Yorke:2002p1}        &  60 & 45 & 33.6 & \nodata\\
        & 120 & 70 & 42.9 & \nodata\\
\hline
        & 100 (A) & $20^+$ & 5.4 (+ 3.4) & $>$ 5.4\\
\protect\citet{Krumholz:2007p1380}        & 100 (B) & $20^+$ & 8.9 (+ 2.4) & $>$
8.9\\ & 200 & $20^+$ & 8.6 (+ 6) & $>$ 8.6 \\
\hline
\protect\citet{Krumholz:2009p10975} & 100 & $75^+$ & 41.5 + 29.2 (+ 28.3) & $>$
70.7\\
\hline
                       &  60 & 939 & 28.2 & 47.0 \\
                       & 120 & 489 & 56.5 & 47.1 \\
% Update tend, M_* and SFE of still running simulation
\raisebox{1.5ex}[-1.5ex]{This study} & 240 & 226 & 92.6 & 38.5 \\
                       & 480 & $41^+$ & 137.2 (+ 67.8) & $>$ 28.5 but $<$ 42.7\\
\enddata 
\tablecomments{
The columns from left to right state 
the authors, 
the initial pre-stellar core mass,
the evolutionary time simulated,
the final star mass,
as well as the corresponding star formation efficiency. 
A ``+'' in the $t_\mathrm{end}$ column means that the accretion phase is not simulated until the end yet.
In the case of the simulation by M.~Krumholz,
only the formation of the most massive stars are considered here;
all other stars formed have masses below 1~$\mbox{M}_\odot$.
In the case of
\protect\citet{Krumholz:2007p1380}
the ``(A)'' and ``(B)'' in the $M_\mathrm{core}$ column mark the usage of different perturbation fields of the initial
state (same labels as in the original paper)
and the $M_*$ column gives additionally the remnant disk mass around the primary star.
In the case of
\protect\citet{Krumholz:2009p10975}
and the 480\Msol case of our own study
the $M_*$ column gives additionally the remnant disk plus envelope mass.
In the case of
\protect\citet{Yorke:2002p1}
only the simulations with frequency dependent radiation transport
are considered.
}
\end{deluxetable} 
The research studies clearly differ in the evolutionary time simulated. 
We improved this by roughly an order of magnitude.
The 20~kyr and 75 kyr of evolution in \citet{Krumholz:2007p1380} and \citet{Krumholz:2009p10975} represent approximately one-third and slightly more than one free fall time of the pre-stellar core respectively.
Despite the frequency dependent radiation transport and the high resolution down to 1.27~AU in our simulations, we are able to follow the evolution of the accreting system up to
% 14.05, 10.46, 6.80, 1.67
% TODO: update the 480Msol accretion/simulation time here
14, 10, 7, and 2 free fall times for an initial core mass of 60, 120, 240, and 480\Msol respectively, including the whole stellar accretion phase.
To state this clearly, this is only possible due to our self-restriction to axial symmetry in these runs.
In the simulations by M.~Krumholz further accretion seems to be the natural continuation of the runs.
Our simulation series of the disk accretion scenario with varying initial core masses shows a decrease of the star formation efficiency towards higher mass cores as a result of the growing radiation pressure feedback.

Simultaneously to the bypass of the thermal radiation by the massive accretion disk, a stable wide-angle bipolar outflow with velocities of the order of $100~\mbox{km~s}^{-1}$ is launched by the radiation pressure.
In these axially symmetric simulations, we did not detect any evidence for a radiative instability of these outflow regions, like observed in the simulations by \citet{Krumholz:2009p10975}.
Following the explanatory notes in \citet{Krumholz:2009p10975}, this is due to the fact, that this instability requires non-axial symmetric modes to occur.
A detailed study of this regime seems to be necessary to clearly understand the underlying physics of this requirement.
On the other hand, our simulations show a strong release of radiation pressure in the bipolar direction, growing in angle with time, which 
\vONE{fits to the observed broadening of outflows}
%vONEgives a reasonable explanation for the observed evolutionary phases of outflows 
in massive star forming regions \citep{Beuther:2005p1805}.
\vONE{But to state this clearly, we are sure that a complete description of the jet or outflow physics cannot be done without taking care of the dominant magnetic field effects.}

\begin{figure}[htbp]
\begin{center}
\includegraphics[width=\FigureWidth]{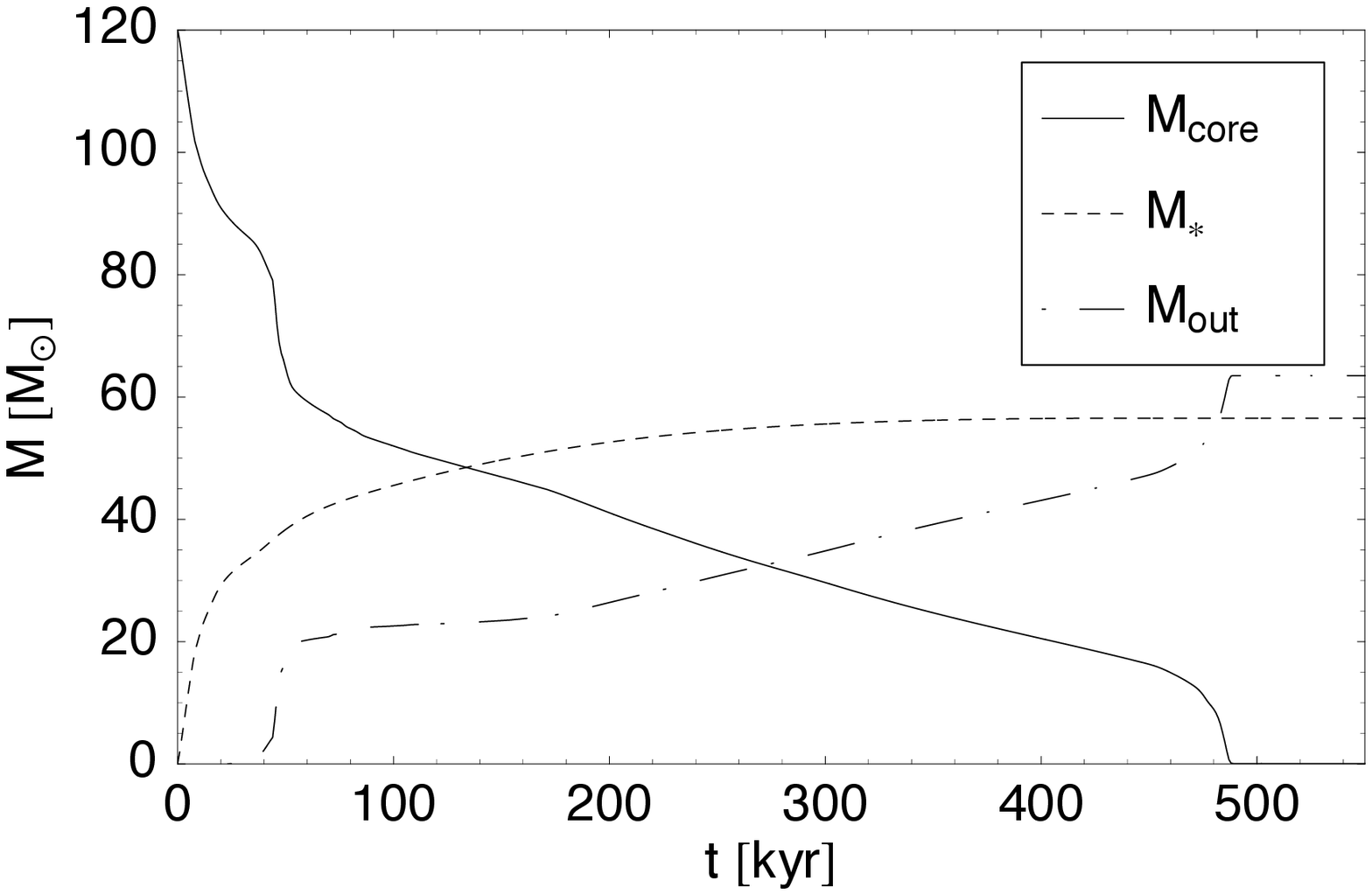}
\caption{
Evolution of mass flows over the inner and outer computational boundary.
\newline
Solid line: Mass inside the computational domain.
\newline
Dashed line: Mass of the forming massive star in the center.
\newline
Dot-dashed line: Mass loss due to the radiation pressure in the radially
outward direction. 
}
\label{fig:2D_Masses}
\end{center}
\end{figure}

Fig.~\ref{fig:2D_Masses} shows the time dependent fractions of masses, divided into the mass inside the computational domain, the mass of the forming star (from the flux over the inner boundary), and the mass loss by the radiation pressure driven outflow (from the flux over the outer boundary).

The different epochs of the collapse of the 120\Msol pre-stellar core are illustrated in Fig.~\ref{fig:SimulationSnapshots}.
The subfigures display the initial condition, the disk formation and evolution, the outflow launching, and the end of the accretion phase in several snapshots of the density structure.
The corresponding animation is available in the online-materials.

\begin{figure}[p]
\begin{center}
\subfloat[Color scale of the gas density,
plotted in logarithmic scale from $10^{-19}$ up to $10^{-15}\rhocgs$.]{
\includegraphics[width=\FigureWidth]{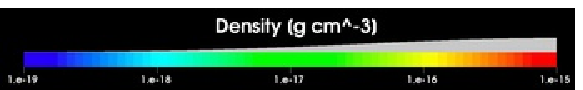}
}
\par
\subfloat[0 kyr]{
\includegraphics[width=0.5\FigureWidth]{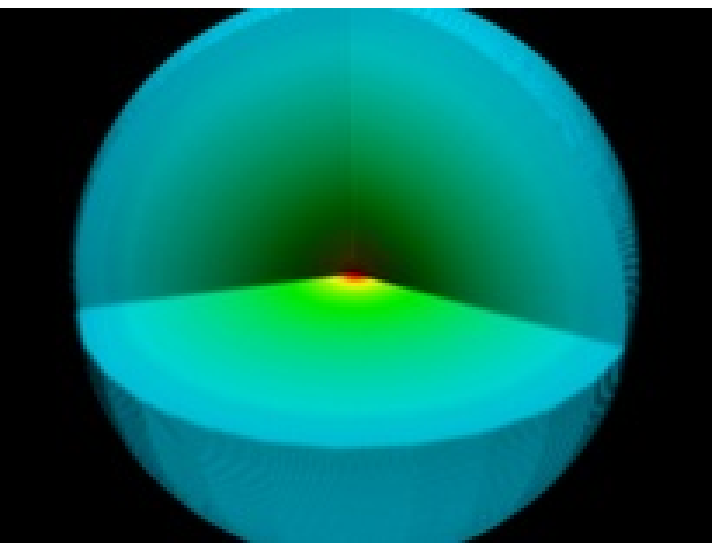}
}
\subfloat[30 kyr]{
\includegraphics[width=0.5\FigureWidth]{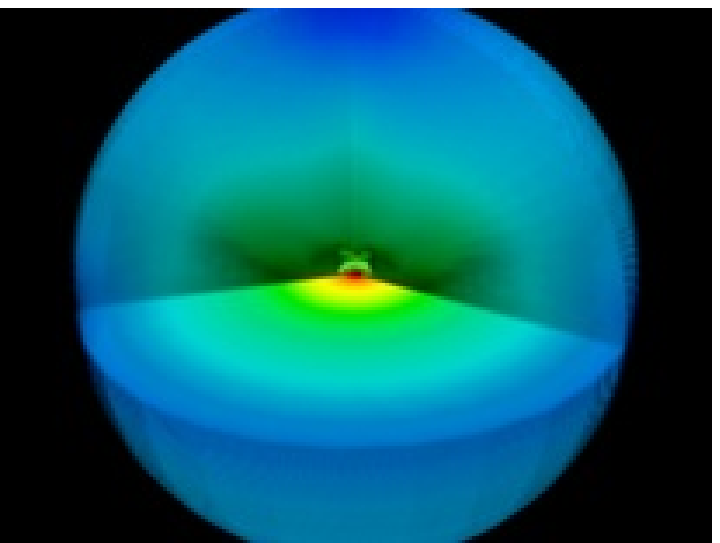}
}
\par
\subfloat[40 kyr]{
\includegraphics[width=0.5\FigureWidth]{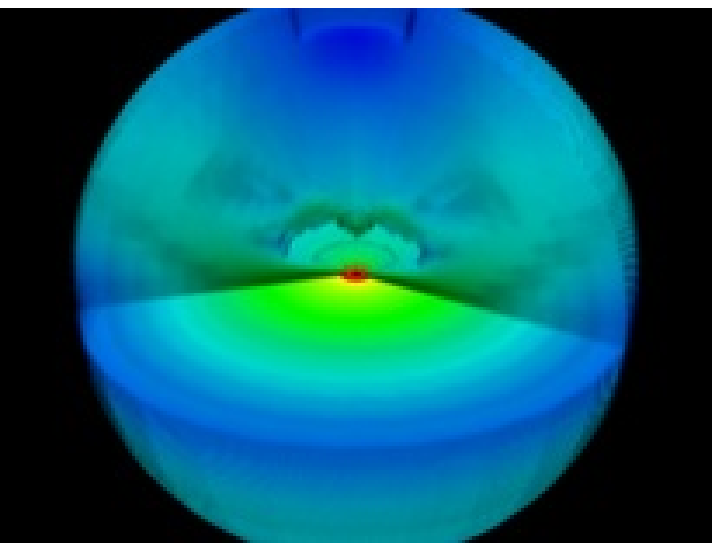}
}
\subfloat[50 kyr]{
\includegraphics[width=0.5\FigureWidth]{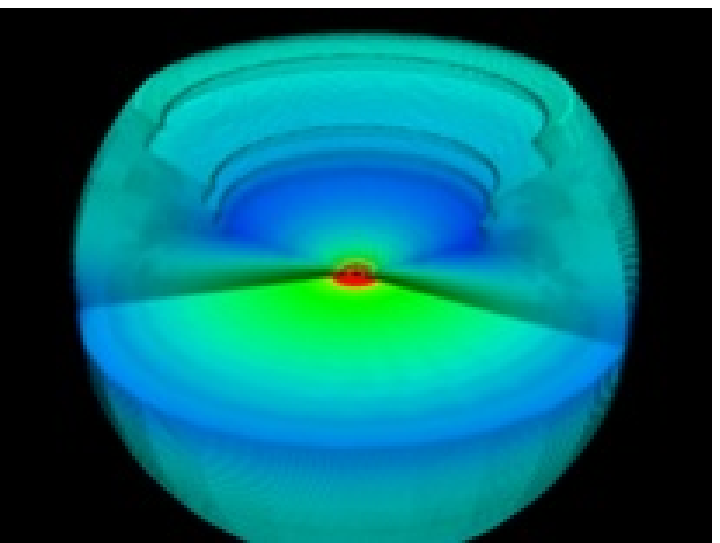}
}
\par
\vspace{\CaptionSpace}
\vspace{-2ex}
\caption{
Simulation snapshots from a collapse of a 120\Msol pre-stellar core. 
All images show the same scale of the whole pre-stellar core with an initial diameter of 0.2~pc.
}
\label{fig:SimulationSnapshots}
\end{center}
\end{figure}

\setcounter{figure}{20}

\begin{figure}[p]
\setcounter{subfigure}{5}
\begin{center}
\subfloat[200 kyr]{
\includegraphics[width=0.5\FigureWidth]{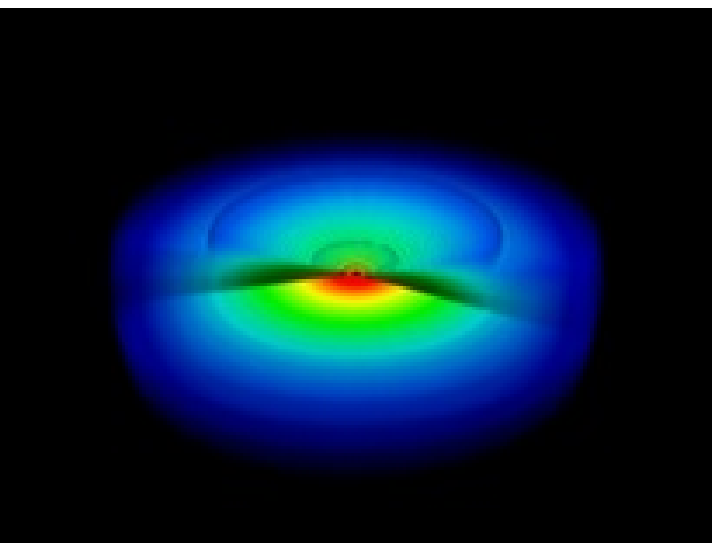}
}
\subfloat[400 kyr]{
\includegraphics[width=0.5\FigureWidth]{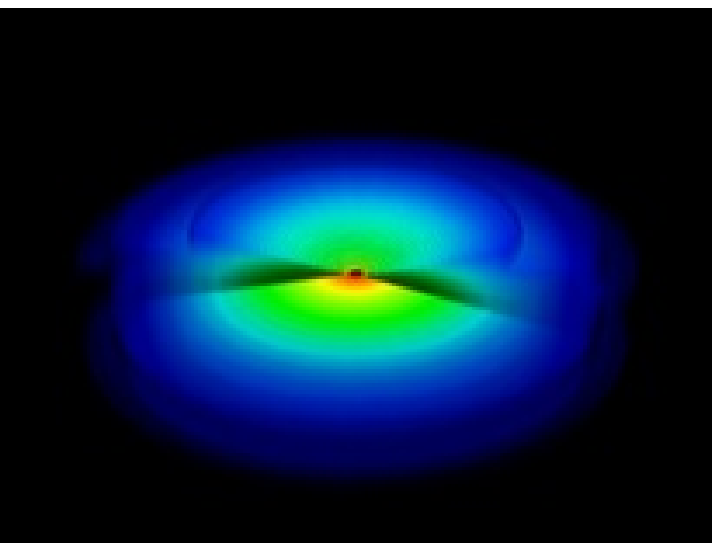}
}
\par
\subfloat[450 kyr]{
\includegraphics[width=0.5\FigureWidth]{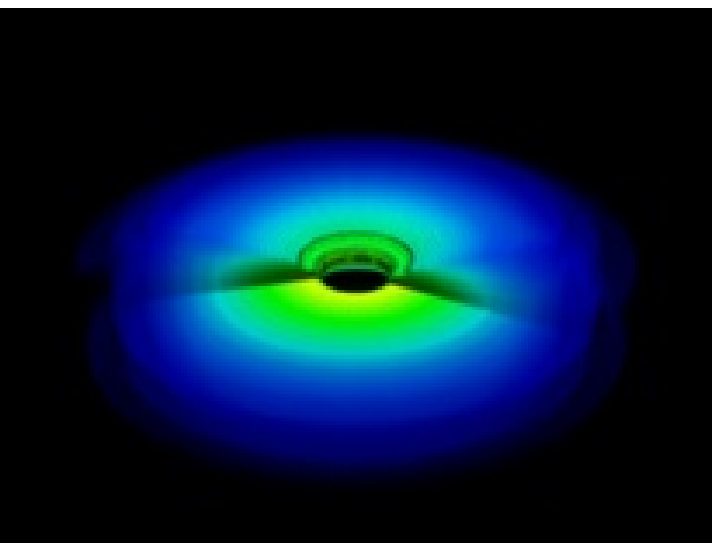}
}
\subfloat[460 kyr]{
\includegraphics[width=0.5\FigureWidth]{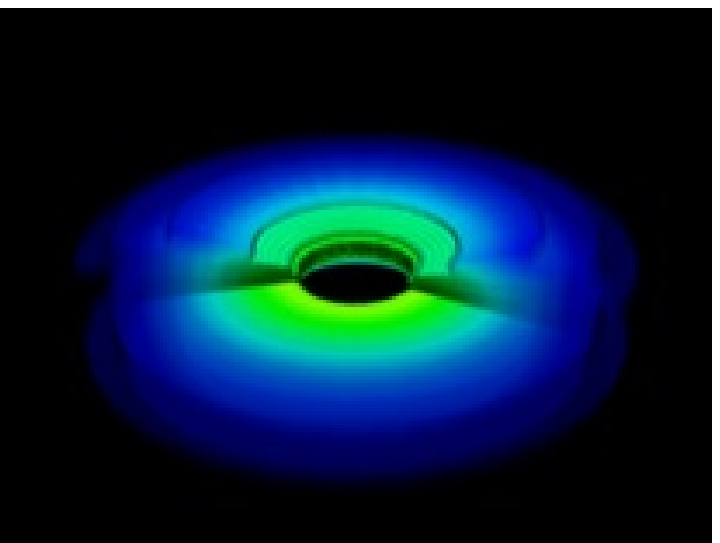}
}
\par
\subfloat[470 kyr]{
\includegraphics[width=0.5\FigureWidth]{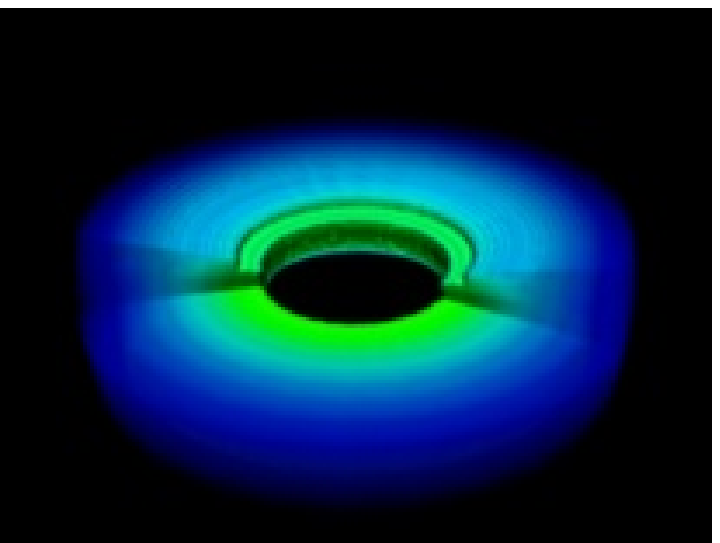}
}
\subfloat[480 kyr]{
\includegraphics[width=0.5\FigureWidth]{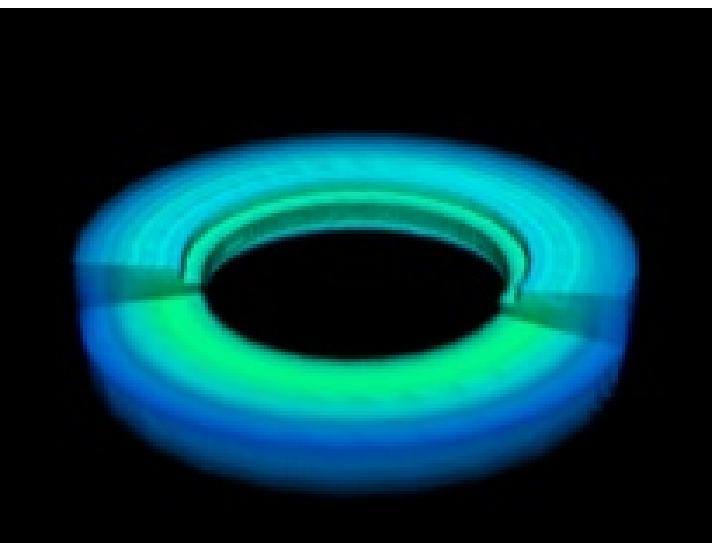}
}
\caption{Continuation of Fig.~\protect\ref{fig:SimulationSnapshots}}
\label{fig:SimulationSnapshotsContinuation}
\end{center}
\end{figure}

\subsection{Limitations of our approach}
A minor flaw of our studies of the radiation hydrodynamics around the dust sublimation front is the usage of a simple constant gas opacity of $\kappa_\mathrm{gas} = 0.01 \mbox{ cm}^2 \mbox{ g}^{-1}$ for the completely evaporated regions around the forming star.
In other words, the neighborhood of the star remains optically thin for the stellar irradiation up to the dust sublimation front. 
From our results of the parameter scan of the inner sink cell radius, we conclude that the prior absorption of the stellar irradiation in dust-free, but potentially optically thick regions, would even enhance the crucial anisotropy of the radiation field detected in our simulations.
In this sense, the ignorance of the detailed optical properties of the gas phase does not imply any loss of generality.

The usage of a central sink cell of a specific radius implies further assumptions, e.g.~we assume that the mass flow into the central sink cell is accreted by the central star.
If, e.g.~the mass flow is transferred into an outflow or jet in this inner region near the stellar surface, the stellar growth would be decreased and the final masses of the stars given here represent upper mass limits.

\vONE{
Due to the fact that higher accretion rates lead first to a delay of the star's approach to the main sequence and secondly to a higher ratio of accretion luminosity to stellar luminosity (which increases the importance of the correct knowledge of the stellar radius),
}
the details of the treatment of stellar evolution become important especially in the case of the highest mass cores studied.
An even more realistic approach than the usage of tabulated tracks could be achieved by including a stellar evolution code such as \citet{Hosokawa:2009p12591} to calculate the ongoing stellar physics in the sink cell consistently in time.

To mimic the effect of angular momentum transport by evolving instabilities in the accretion disk, we made use of the $\alpha$-viscosity model by \citet{Shakura:1973p3060}.
Nevertheless, all our code development (radiation transport and self-gravity) and model setup is already in three-dimensional formulation.
Hence we started to expand our simulations into full 3D.
First, this will allow to compute the angular momentum transport (by gravitational torques) consistently with the formation and evolution of the accretion disk.
Secondly, 
\vONE{although we will not study the fragmentation of the outer core regions,}
the stability or potential fragmentation of the forming massive circumstellar disk can be addressed.

\section{Summary}
\label{sect:conclusion}
We performed 
high-resolution 
radiation hydrodynamics simulations 
of monolithic pre-stellar core collapses 
including frequency dependent radiative feedback.
A broad parameter space of
various numerical configurations and initial conditions
was scanned.
The dust sublimation front in the vicinity of the forming star could be resolved down to 1.27~AU.
The evolution of the system was computed over its whole accretion phase of several $10^5$~yr.
The usage of frequency dependent ray-tracing in our newly developed radiation module denotes the most realistic
radiation transport method used in multi-dimensional hydrodynamic simulations of massive star formation by now. 
The broad parameter studies, 
especially regarding the size of the sink cell and the initial core mass,
reveal new insights of the radiative feedback onto the accretion flow
during the formation of a massive star:

In the case of spherically symmetric accretion flows,
we confirm the results of previous research studies
\citep{Larson:1971p1210, Kahn:1974p1200, Yorke:1977p1358, Wolfire:1987p539} 
that the thermal radiation pressure by re-emitted photons 
behind the dust sublimation front
overcomes gravity,
stops the accretion flow,
and finally reverts the in-falling envelope.
The upper mass limit of spherically symmetric accretion for our specific dust 
\citep{Laor:1993p736}
and stellar evolution model
\citep{Hosokawa:2009p12591}
constrains the final stellar mass to be less than 40$\mbox{ M}_\odot$.

In the case of disk accretion,
the thermal radiation field generates a strong anisotropic feature,
similar to the flashlight effect,
which focus lies on the escape of radiation through optically thin cavities.
We found that it is strict necessary to include the dust sublimation front in the computational domain,
to reveal the persistent anisotropy during the long-term evolution of the accretion disk.
This requirement as well as a steady feeding of the accretion disk from the outer core regions
maintain the anisotropic structure of the thermal radiation field.
The short accretion phases of the disks in the simulations by \citet{Yorke:2002p1} are a result of the fact
that they did not include the dust sublimation front in their simulations,
as clearly shown in our result of the parameter scan of the size of the central sink cell
(see Sect.~\ref{sect:2Drmin}).
Additional feeding of the disk by unstable outflow regions, 
as stated in \citet{Krumholz:2009p10975},
would enhance this anisotropy
but is not necessary.
As a consequence,
we conclude that the radiation pressure problem in the formation of massive stars can be reduced to the question 
if the non-spherically symmetric stellar environment is dense or opaque enough to generate a strong anisotropy of
the thermal radiation field. 

These mechanisms allow 
the central star to increase its mass far beyond the upper mass limit found in
the case of spherical accretion flows. 
For an initial mass of the pre-stellar host core of
60, 120, 240, and 480\Msol
the masses of the final stars formed in our simulations of the disk accretion
scenario add up to
% TODO: update no.
28.2, 56.5, 92.6, and at least 137.2\Msol respectively.
Indeed, 
the final massive stars are the most massive stars ever formed in a multi-dimensional radiation hydrodynamics
simulation so far.

\acknowledgments
This research has been supported by the 
International Max-Planck Research School for Astronomy and Cosmic Physics at the University of Heidelberg (IMPRS-HD).
Author H.~Klahr has been supported in part by the 
Deutsche Forschungsgemeinschaft (DFG) through 
grant DFG Forschergruppe 759
``The Formation of Planets: The Critical First Growth Phase''.
Author H.~Beuther acknowledges financial support by the Emmy-Noether-Programm of the 
DFG through grant BE2578.
Many of our simulations
scanning the broad parameter space 
have been performed on the rio as well as on the pia cluster of the Max Planck computing center in Garching.
We acknowledge
Andrea Mignone, the main developer of the open source magneto-hydrodynamics code Pluto,
as well as 
Petros Tzeferacos, 
who implemented the viscosity tensor into Pluto.
Author R.~Kuiper thanks especially
Cornelis P.~Dullemond, Mario Flock, Johannes Sch\"onke, and Takashi Hosokawa
for their contributions.
Furthermore,
R.~Kuiper thank 
Harold Yorke, Neal Turner, J\"urgen Steinacker, Benoit Commercon, Bhargav Vaidya, and Cassandra Fallscheer
for fruitful discussions.

\appendix

%ONE\vONE{
%ONE\section{The Poisson solver}
%ONEThe discretization of Eq.~\eqref{eq:Poisson} yields the vector equation
%ONE\begin{equation}
%ONE\label{eq:Poisson_discrete}
%ONEA \vec{\Phi} = 4 \pi ~ G ~ \vec{\rho}.
%ONE\end{equation}
%ONEFor a one-dimensional cartesian grid with a uniform grid spacing $\Delta x$,
%ONEwhere the $i^\mathrm{th}$ component of the vectors $\vec{\Phi}$ and $\vec{\rho}$ represent the gravitational potential
%ONEand density of the $i^{th}$ grid point,
%ONEthe matrix $A$ would be of the form
%ONE\begin{equation}
%ONEA = \left(
%ONE\begin{array}{r r r r r r r}
%ONE\ldots & \ldots & \ldots & \ldots & \ldots & \ldots & \ldots \\
%ONE\ldots & 1 & -2 & 1 & 0 & \ldots & \ldots\\
%ONE\ldots & 0 & 1 & -2 & 1 & 0 & \ldots \\
%ONE\ldots & \ldots & \ldots & \ldots & \ldots & \ldots & \ldots
%ONE\end{array}
%ONE\right)
%ONE\frac{1}{(\Delta x)^2},
%ONE\end{equation}
%ONErepresenting the discretization stencil 
%ONE$(\Phi_{i-1} - 2 \Phi_i + \Phi_{i+1})/(\Delta x)^2$.
%ONEThe desired approximate matrix inversion for solving 
%ONEEq.~\eqref{eq:Poisson_discrete} 
%ONEis done using the so-called GMRES method.
%ONEwhich is also used for the FLD Eq.~\eqref{eq:rad_diffusion}.
%ONEThe accuracy of the Poisson solver,
%ONEi.e.~the abort criterion for the approximate matrix inversion,
%ONEis choosen to 0.001\% relative accuracy of the gravitational potential 
%ONE$\left(\Delta \Phi_\mathrm{sg} / \Phi_\mathrm{sg} \le 10^{-5}\right)$.
%ONE}

\section{Parameter scans of the resolution}
Numerical hydrodynamics simulations involve a discretization of the underlying equations of hydrodynamics
given in continuous space,
cp.~Eqs.~\eqref{eq:hydrodynamics_density}-\eqref{eq:hydrodynamics_energy}. 
This causes a discretization error, 
which in general vanishes for infinitely high resolution of the numerical solver method. 
To compute a specific quantity, 
such as the accretion history, 
with a specific accuracy therefore needs a specific resolution, 
which is necessary to damp the discretization errors down to the requested accuracy. 
In this way it is possible to guarantee the achievement of a converged result.
Although this procedure is a must in numerical research to achieve reliable results, 
the overhead of cpu time needed for convergence runs inhibits their realization in most present-day astronomical
simulations,
especially in multi-dimensional radiation hydrodynamics,
which are performed almost at the upper limit of the computational power of the available clusters.
To fix the number of grid cells, 
which are necessary for a correct representation of the radiation fluid interactions,
we perform several simulations with varying resolution. 
Focusing on the inner regions of the pre-stellar core, 
the radial cell sizes of the grid thereby grow logarithmically from inside out as described in
Sect.~\ref{sect:discretization}. 

\subsection{1D convergence runs}
\label{sect:1DConvergence}

\subsubsection{Simulations}
\begin{figure}[htbp]
\begin{center}
\includegraphics[width=\FigureWidth]{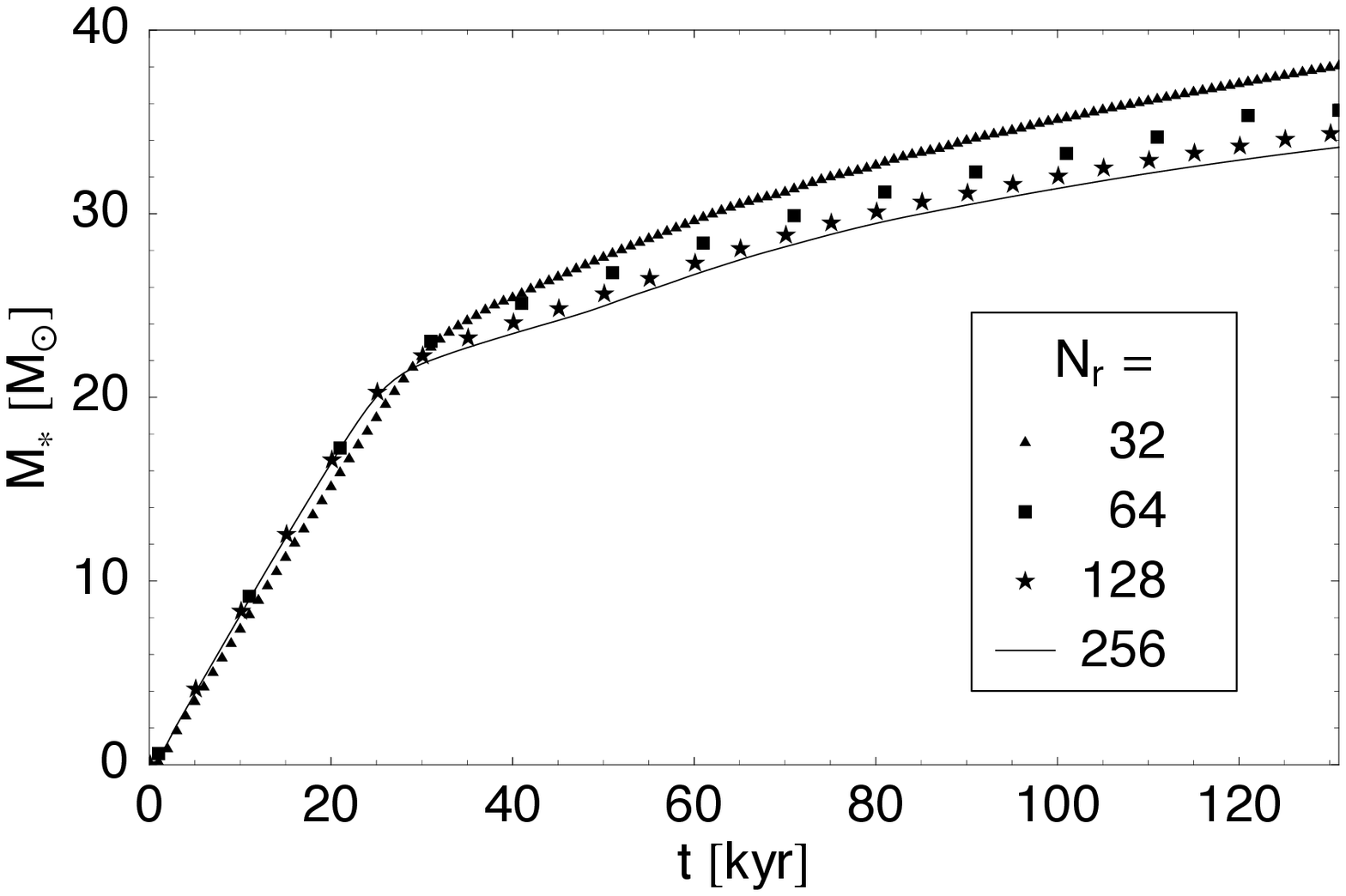}
\caption{
Stellar mass $M_*$ as a function of time $t$ for four different resolutions of the spherically symmetric pre-stellar
core collapse simulations. 
The number of grid cells $N_r$ varies from 32 to 256, 
corresponding to a size of the smallest grid cell of 
$(\Delta r)_\mathrm{min} =$~0.36~AU to 0.04~AU respectively.
}
\label{1D_Convergence_Rmin1AU}
\end{center}
\end{figure}

The initial conditions and numerical parameters of these convergence runs are described in
Sect.~\ref{sect:InitialConditions}. 
The simulations are performed for an initial core mass of 
$M_\mathrm{core} = 60 \mbox{ M}_\odot$ and with an inner sink
cell radius of $r_\mathrm{min} = 1$~AU.
We follow the long-term evolution of the system for at least 
%TODO update Nr=256 run and no. accordingly
163 kyr, 
representing 2.4
free fall times of the pre-stellar core. 
The resulting mass growth $M_*(t)$ of the centrally forming star is displayed in 
Fig.~\ref{1D_Convergence_Rmin1AU}. 

\subsubsection{Conclusions}
The lowest resolution run ($N_r = 32$) fails to compute the correct amount of
accretion already during the mostly isothermal initial free fall phase (up to 25~kyr).
For higher resolution runs with 64 grid cells or more, 
the mass growth of the forming star is identical during this phase. 
At a later evolutionary epoch, 
when radiative feedback becomes important, 
simulations with higher resolution lead generally to a slower mass growth.
The deviation of a specific run to the next one with double resolution shrinks towards higher resolution,
so the simulations fulfill the requirement of a monotonic convergence.
Our one-dimensional simulations with varying initial core masses
(Sect.~\ref{sect:1DMcore}) use 128 grid cells in the radial direction and an inner radial boundary at $r_\mathrm{min} = 1$ AU corresponding to a grid size of
$(\Delta r)_\mathrm{min} = 0.08$ AU for the innermost grid cell.

\subsection{2D convergence runs}
\label{sect:2DConvergence}
\subsubsection{Simulations}
\begin{figure}[p]
\begin{center}
\includegraphics[width=\FigureWidth]{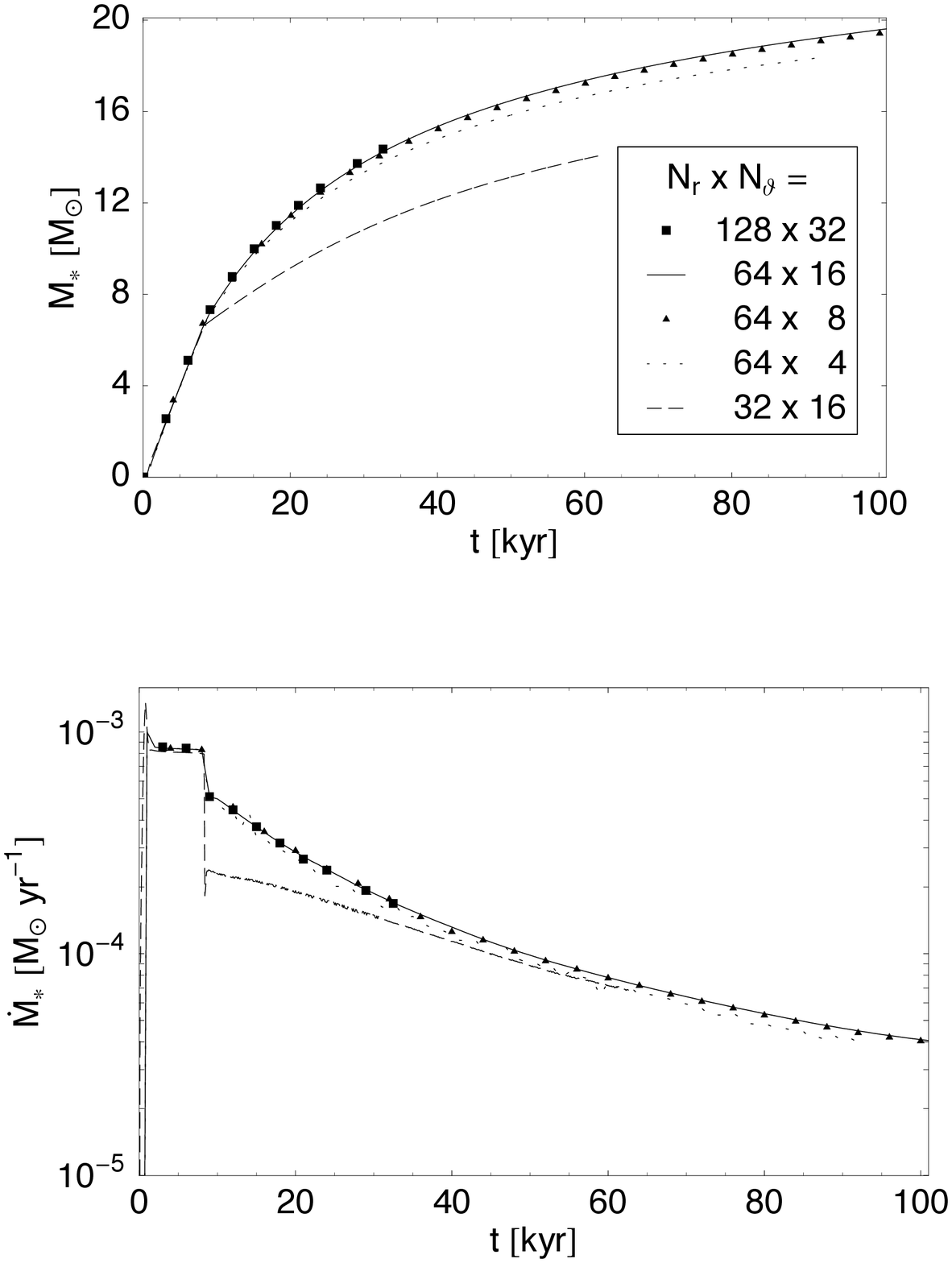}
\caption{
Stellar mass $M_*$ (upper panel) and accretion rate $\dot{M}_*$ (lower panel) as a function of time $t$ for five
different resolution to determine the adequate number of grid cells necessary for the collapse simulations of
the rotating pre-stellar cores. 
}
\label{2D_Convergence}
\end{center}
\end{figure}

\begin{figure}[p]
\begin{center}
\includegraphics[width=\FigureWidth]{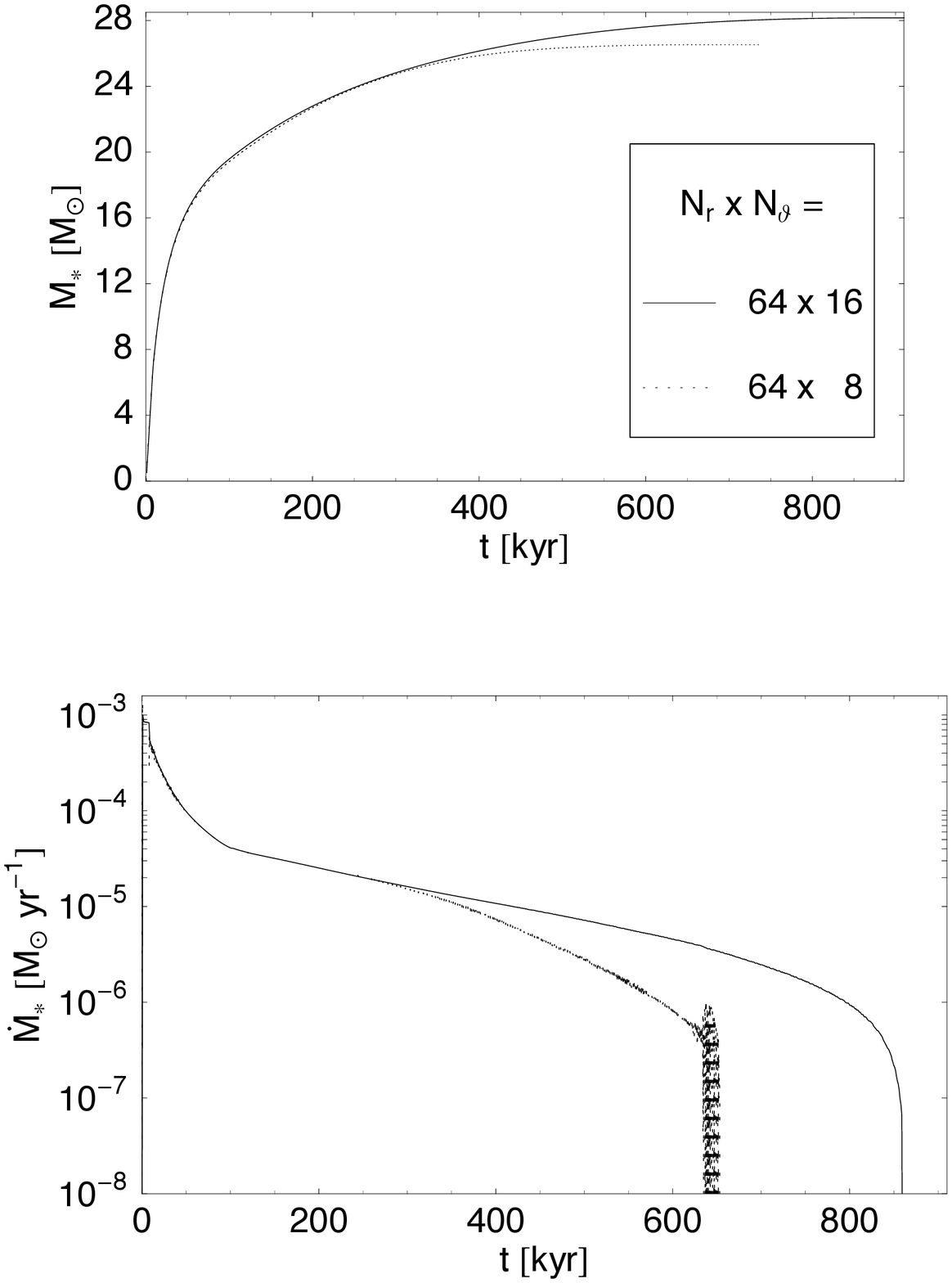}
\caption{
Stellar mass $M_*$ (upper panel) and accretion rate $\dot{M}_*$ (lower panel)
as a function of time $t$ for two different resolution in a 
long-term convergence study up to the end of the disk accretion phase.
}
\label{2D_Convergence_long-term}
\end{center}
\end{figure}

To fix the number of grid cells necessary for computing the correct physics of the radiation fluid interaction, 
we performed several simulations with varying resolution in the two-dimensional
setup, too. 
The basic initial conditions and numerical parameter used for these
convergence runs are described in Sect.~\ref{sect:InitialConditions}. 
The simulations are performed for a core mass of 
$M_\mathrm{core} = 60 \mbox{ M}_\odot$
and the inner boundary of the computational domain is located at
$r_\mathrm{min} = 10$~AU. 
We followed the evolution of the collapsing core
up to 33 kyr (0.5 free fall times) for the highest resolution case yet 
and up to several hundred kyr 
(about 10 free fall times) for a long-term convergence run.
The accretion history and the corresponding mass growth of the centrally forming star are displayed in
Figs.~\ref{2D_Convergence} and \ref{2D_Convergence_long-term}.

\subsubsection{Conclusions}
In contrast to the purely one-dimensional in-fall (Sect.~\ref{sect:1DConvergence}), 
the centrifugal forces slow down the radially proceeding dynamics.
So the usage of 64 grid cells in the radial direction, 
corresponding to a radial grid size of the innermost cells of 
$(\Delta r)_\mathrm{min} = 1.27$~AU, 
yields a converged result for the accretion rate onto the forming high-mass star.
The low-resolution run with only 32 grid cells in the radial direction clearly fails to compute the correct onset of
disk formation after 8~kyr. 
Due to the clear dominance of the motion of gas in the radial direction during the initial `free fall' phase up to
roughly 8~kyr the accretion rates of this epoch are independent of the resolution used in the
polar direction. 
The required resolution in the polar direction to compute a converged result also during later epochs remains notably
poor, 
reflecting the fact that the complex radiation hydrodynamics interactions act mostly in the radial direction.
This result confirms the expedient choice of spherical coordinates in monolithic core collapse simulations.
Higher resolution of the polar stratification of the forming circumstellar disk mostly influences the cooling of the
irradiated and viscously heated midplane layer.
The usage of only 4 or 8 grid cells in the polar direction therefore results in stronger fluctuations of the accretion
flow, 
which vanish in the higher resolution runs 
(clearly visible in the lower panel of Fig.~\ref{2D_Convergence}).
On the other hand the runs with low resolution in the polar direction underestimate the mass growth of the forming
star only slightly 
(upper panel in Fig.~\ref{2D_Convergence}).
The deviations of each run to the next run in higher resolution shrink towards higher resolution, 
that means the simulation series yields a monotonous convergence.
The long-term convergence study (Fig.~\ref{2D_Convergence_long-term})
clearly shows that the point in time when the disk looses its shadowing property
depends on the polar resolution of the circumstellar disk.
Higher resolution of the disk's stratification results in a stronger anisotropy of the
thermal radiation field and therefore minimizes the radiation pressure on the
accretion flow.
%\vONE{
%The polar resolution used in our simulations satisfies the requirement that a single grid cell is equal or smaller than the pressure scale height of the disk, which represents the typical lengthscale of the cooling property of the disk, e.g.~$\Delta x =$~2-3~AU at the dust sublimation front at $r =$~20-30~AU.
%}

The runs with 64~x~16 and 128~x~32 grid cells show fully converged results even during the epoch
of the most rapid changes at the onset of disk formation at 8~kyr. 
The spike in the accretion rate downwards during this onset represents the short period in time, in which for the first
time a fluid package from the outer core region arrived at the innermost radius $r_\mathrm{min}$ with high enough
angular momentum to compensate the stellar gravity.
Quickly hereafter the following mass builds up a circumstellar disk, in which the shear viscosity yields an
angular momentum transfer outwards resulting in a steady accretion rate anew.
At later evolutionary phases the amplitude of the accretion rate is mostly a result of a quasi-stationary accretion
flow inwards and an interactive radiative flux in the outward direction, which smoothly grows proportional to the
luminosity of the forming massive star.
The deviations of the individual runs during this more evolved and `less violent' phase shrink again for all
resolutions studied. 
Our two-dimensional simulations presented use 64 x 16 grid cells as the default
setup corresponding to a grid size of $(\Delta r \mbox{ x } r \Delta \theta)_\mathrm{min}$ = 1.27 AU x 1.04 AU for the innermost grid
cells.

% apj.bst downloaded from www.astro.virginia.edu/coolflow/apj.bst
\bibliographystyle{apj}
\bibliography{Papers}
\end{document}